\documentclass{emulateapj}
\usepackage{apjfonts}
\usepackage{hyperref}
\usepackage{epsf}

\shortauthors{Gy\H{o}ry \& Bell}
\shorttitle{The merger origin of early-type galaxies}

\slugcomment{{\sc The Astrophysical Journal, in press} }
\newcommand{\etal}{{et~al.}}

\begin{document}



\title{Testing a prediction of the merger origin of early-type galaxies: 
a correlation between stellar populations and asymmetry}

\author{Zsuzsanna Gy\H{o}ry$^{1,2}$ and Eric F.\ Bell$^{3,2}$ }
\affil{$^1$ Department of Physics, E{\"o}tv{\"o}s University,
Budapest, Pf.\ 32, H-1518 Hungary; \texttt{gyory.zsuzsa@googlemail.com} \\
$^2$ Max-Planck-Institut f\"ur Astronomie,
K{\"o}nigstuhl 17, D-69117 Heidelberg, Germany   \\
$^3$ Department of Astronomy, University of Michigan, 500 Church St., Ann Arbor, MI 48109, USA; \texttt{ericbell@umich.edu} }

\begin{abstract}
One of the key predictions of the merger
hypothesis for the origin of early-type (elliptical and lenticular) 
galaxies is that 
tidally-induced asymmetric structure should correlate 
with signatures of a relatively 
young stellar population.  Such a signature was found 
by Schweizer and Seitzer (1992;\,AJ,\,104,\,1039) at roughly 
4$\sigma$ confidence.  In this paper, 
we revisit this issue with a nearly ten-fold larger 
sample of $0.01<z<0.03$ galaxies selected from 
the Two Micron All-Sky Survey and the Sloan Digital Sky Survey.
We parameterize tidal structure using a repeatable 
algorithmic measure of asymmetry, and correlate this
with color offset from the early-type galaxy color--magnitude
relation.  We recover the color offset--asymmetry correlation; 
furthermore,  
we demonstrate observationally for the 
first time that this effect is driven by a highly-significant trend
towards younger ages at higher asymmetry values.  
We present a simple model for the evolution
of early-type galaxies through gas-rich major and minor mergers
that reproduces their observed 
build-up from $z=1$ to the present day
and the distribution of present-day colors and ages.  
We show using this model that if both stellar populations and asymmetry were 
ideal `clocks' measuring the time since last major or minor gas-rich 
interaction, then
we would expect a rather tight correlation between age
and asymmetry. We suggest that the source of extra 
scatter is 
natural diversity in progenitor star formation history, gas content, 
and merger mass ratio, but quantitative
confirmation of this conjecture will require sophisticated 
modeling.  
We conclude that the asymmetry--age correlation is in basic
accord with the merger hypothesis, and indicates 
that an important fraction of the early-type galaxy population 
is affected by major or minor mergers at cosmologically-recent times.
\end{abstract}



\section{Introduction}
\label{sec:intro}

It has become clear in the last decade that the distribution of galaxies
is bimodal in color, and that this bimodality correlates
with the structure (or morphological type) of galaxies 
\citep{strateva01,kauffmann03,blanton03}. 
Early-type galaxies are dominated by a spheroid of stars
supported largely by random motions, and tend to have red colors and
very low present-day star formation rates.  They populate
a tight color--magnitude relation (CMR; \citealp{sandage78}, 
\citealp{bower92}, \citealp{ruhland07}); 
the tightness of this CMR is an indicator
of their rather ancient stellar populations \citep{trager00,gallazzi06}.
In contrast, late-type
galaxies are dominated by a rotationally-supported disk of 
stars, and tend to have significant ongoing star formation at
the present day, and populate a reasonably
broad (at optical wavelengths) blue cloud whose colors reflect a combination 
of recent star formation and dust \citep{sch08}.
The stellar mass functions for the two classes of 
galaxy are different; early-type galaxies can extend to much higher stellar
mass than late-type galaxies \citep{bell03}.  
This bimodality persists at all redshifts up to $z \sim 1$
\citep{bell04,bell_gems,faber07}.

One of the most remarkable features of the bimodality in the 
galaxy population is the observed correspondence between 
galaxy structure and stellar populations.  Models
of galaxy formation in a cosmological context predict 
that the structure of a galaxy reflects the dynamical 
assembly (merger history) of a galaxy \citep{kauf93,cole00}, whereas
the cooling of gas and subsequent star formation is a 
property of the warm/hot reservoir of gas in a halo, which is
relatively unaffected by galaxy interactions \citep{cole00,cattaneo06}.
Galaxy formation models with `standard' ingredients --- the 
formation and evolution of dark matter halos, 
gas cooling, star formation and stellar feedback --- 
predict that almost all galaxies should be forming 
stars at an appreciable rate \citep{cole00,croton06,cattaneo06,keres09}.
This is in direct disagreement with observations of a red sequence, 
and is usually interpreted (for good reasons) as indicating not just 
a minor failing of (admittedly uncertain) prescriptions for 
these physical processes; rather, this deficiency is 
interpreted as the signature of a completely separate
physical mechanism.

In this regard, the empirical correlation between 
a lack of star formation and a prominent stellar spheroid
(bulge) is an important clue to the mechanism suppressing 
star formation \citep{kauffmann06}.  A bulge appears to be 
a necessary (but not sufficient) requirement for 
the shut-down of star formation \citep{bell08}.  Taken together
with the bulge mass--black hole mass correlation 
\citep{mag98,haering04}, it is not unreasonable to postulate
that feedback from accretion onto a supermassive black 
hole (active galactic nucleus [AGN] feedback) is an important mechanism by which galaxies
quench their star formation 
\citep{dimatteo05,springel05,croton06,bower06,cattaneo06,hopkins08a,hopkins08b,somer08,johansson09feedback}; 
although other physical mechanisms may also play
a role (see, e.g., \citealp{naab07}, \citealp{db07}, \citealp{khochfar07}, 
\citealp{johansson09grav}, 
\citealp{dekel06}, \citealp{birnboim07} or
\citealp{guo07} on gravitational heating, the influence of the development 
of virial shocks, and the 
heating of large halos with cosmic ray energy).

Galaxy merging plays a decisive role in this picture.
\citet{toomre} and \citet{barnes92} argued that a natural formation route
for early-type galaxies (in particular elliptical galaxies) was 
through the major merger of two pre-existing galaxies (although 
see, e.g., \citealp{burkert08}, \citealp{naab09} 
for discussions of limitations of this obviously
over-simplified picture).
In the intervening time, an impressive variety of evidence
has been amassed showing that the properties of early-type 
galaxies, at least at the broad level, are consistent with those 
expected for merger remnants: e.g., overall kinematics \citep{naab06kin,jesseit07,hoffman09}, 
the widespread existence of
kinematically-decoupled cores \citep{emsellem07}, the distribution of
isophotal shapes \citep{naab03,naab06}, surface brightness profile
shapes \citep{naab06sb,hopkins08sb}, and
low-level tidal 
debris around many elliptical galaxies \citep{malin,ss92,vd05,tal09}.  
Observations of merger remnants
support this picture also, having surface brightness
profiles, velocity dispersions and sizes in the near-infrared 
that are similar to intermediate-luminosity early-type galaxies
\citep{dasyra06,rothberg06}.  

It is argued that shocks and non-circular motions in a 
galaxy merger cause the gas to lose angular momentum,
leading to the growth of the black hole mass and 
conditions suitable for efficient coupling of energy 
from an AGN to the diffuse gas in a galaxy
\citep{sanders,hopkins08a,hopkins08b}.  Circumstantial support
for this picture of merger-driven AGN activity and feedback is 
seen in the tendency for the AGN hosts to have intermediate
colors (between blue cloud and red sequence; \citealp{schawinski07,schawinski10}), rapidly-outflowing metal-enriched gas at $\sim 100$kpc from an active 
quasar at $z\sim 2.4$ \citep{prochaska09}, 
and rapidly-outflowing gas ($\ga 1000$\,km\,s$^{-1}$) in post-starburst
galaxies --- such velocities are not expected to result from 
starburst-driven winds and were interpreted as being relics of 
a quasar-driven wind \citep{tremonti08}.    In such 
a picture, the suppression (or quenching) of star formation 
is reasonably rapid and follows the merger event \citep{kauf00,dimatteo05}.

Another aspect that has come into focus in the last years
is that the formation of red sequence (and early-type) galaxies
is an ongoing process; in particular, more than half of the 
early-type galaxy population has come into place 
since $z \sim 1$ (see, e.g., \citealp{bell04}, \citealp{bell_gems}, \citealp{faber07}, \citealp{brown07}; see \citealp{vd10} and \citealp{robaina10} for analyses focused on the most massive galaxies).  
Thus, one expects that the mergers (that create the early-type
galaxies) and the physical processes that lead to the truncation 
of star formation (that make those early-types red) continue
to happen at relatively recent times.  In this case, these
physical processes, and their late-time signatures, 
should be observable at the present day (see \citealp{hopkins10}
for a detailed and thoughtful discussion of merging and early-type galaxy 
evolution in a cosmological context).

In this context, the work of \citet{ss92} is of much 
importance.  
Motivated by the merger hypothesis for the 
origin of early-type galaxies, they sought a correlation 
between fine structure (from a tidal origin) and 
color offset from the CMR (in \citealp{s90} they did a similar 
work with a smaller sample with absorption line indices).
If gas-rich mergers are the primary way in which early-type
galaxies are made, then one expects a correlation between 
color offset and fine structure. \citet{ss92} found such a correlation
with $\sim 4 \sigma$ confidence, 
albeit with considerable scatter; \citet{tal09} found a similar 
result with a complete sample of nearby ellipticals.   \citet{ss92} used 
stellar population models to explore this result in more 
detail, finding that the early-type galaxy colors 
could be reproduced by a variety of post-merger ages
from 2 to 6 Gyr, depending on the model parameters
used.  

Yet, a number of issues are left open by these works.
Number statistics was a clear issue in both cases (e.g., \citealp{ss92}
 had a total of 69 galaxies in their sample; \citealp{tal09} 
had 55 galaxies in their sample), and testing with a larger 
sample is clearly desirable.  
The definition of fine structure adopted by \citet{ss92}
is motivated by the type of tidal features regularly seen around
nearby early-type galaxies \citep{malin,ss88,vd05,tal09}, but
has two limitations.  Firstly, it is impossible to measure their 
fine structure parameter automatically. \citet{tal09} partially 
remedied this by defining a reproducible metric that is readily applicable
to deep data, although they studied a sample of
elliptical galaxies only.  Secondly, 
high-contrast tidal features are often
the result of {\it minor} mergers or accretions, not 
major mergers.  Finally, at the time at which 
\citet{ss92} was written, it was impossible to disentangle age 
and metallicity influences on spectral indices or colors; it is 
important to verify that age is in fact the driving parameter
of any correlations seen.  

Cognizant of these issues, we have initiated an effort
to confirm and expand on the landmark results of \citet{ss92}.
We have defined a volume and $K$-band luminosity limited 
sample (\S \ref{sec:data}) of $0.01<z<0.03$ early-type galaxies using the Two Micron All-Sky Survey
(2MASS; \citealp{2mass}), the Sloan Digital Sky Survey (SDSS; \citealp{dr3}), 
supplemented with redshifts from the NASA/IPAC Extragalactic Database
(NED).  We have used asymmetry as an automated and repeatable 
metric for tidally-induced structure, and have analyzed it in 
conjunction with SDSS colors, and, crucially, ages and metallicities 
derived from absorption-line spectroscopy \citep{gallazzi05}; these 
parameters are described in \S \ref{sec:color-asymm}.  Because of our 
use of the SDSS as our imaging data we analyze the brighter inner parts
of early-type galaxies (less sensitive diagnostic of faint tidal 
features) but with the advantage of a larger sample (over 600 galaxies).
We present our results in \S \ref{sec:results}.
Finally, 
bearing in mind the insight gleaned from redshift
surveys of red sequence and early-type galaxy 
evolution, we were in a position to model the 
spectral properties of early-type galaxies more completely than 
\citet{ss92}; this discussion is presented in \S \ref{sec:disc}.
The casual reader is invited to skip to \S \ref{sec:results} 
directly, 
focusing in particular on Figs.\ \ref{fig:asymm-color} and 
\ref{fig:asymm-age}. 
In what follows, we use AB magnitudes and assume 
H$_0 = 70\,{\rm km\,s^{-1}\,Mpc^{-1}}$, $\Omega_{m,0} = 0.3$ and
$\Omega_{\Lambda,0} = 0.7$. 

\section{Data and fitting}
\label{sec:data}

\subsection{Sample selection}
\label{sec:sel}

Our aim is to test the relationship between fine structure and color
deviation from the CMR for early-type galaxies.
In order to test this relationship, a representative (but not necessarily
complete) sample is required, and at distances close enough to allow 
us to discern both the global and tidal structure of early-type galaxies.
Accordingly, we have chosen galaxies within a relatively
thin, and nearby, distance shell $0.01<z<0.03$ for study.  The bulk of our 
sample can be drawn from the SDSS, but the relatively 
bright galaxies in this sample lack SDSS spectra (for a variety of reasons, but partially because their fiber magnitudes
were too bright to target as part of the SDSS main survey).  

Accordingly, we have taken a hybrid approach for this work.
The primary object selection was made from the 2MASS All-Sky Extended
Source Catalog. We selected galaxies brighter than 13.5 mag in
$K$. We retrieved redshifts for these
objects either from NED\footnote{The NASA/IPAC Extragalactic Database, 
\url{http://nedwww.ipac.caltech.edu/}} or SDSS,
 matching the objects by position; 
NED was used to fill in objects lacking spectroscopy from the SDSS.
 In order to achieve a good redshift completeness
 we only included galaxies from regions
 of the sky that are fully covered by spectroscopic sample of the SDSS
 DR3. 

\begin{figure}[t]
\begin{center}
\plotone{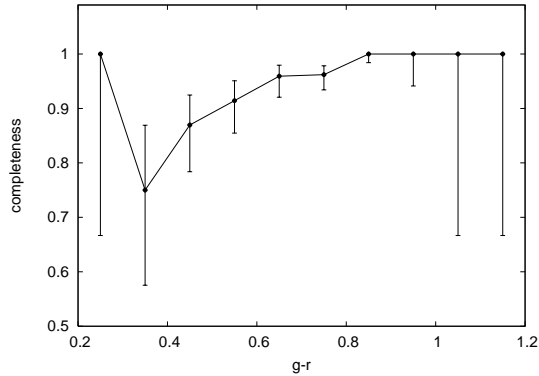}
\caption{The estimated completeness of a 2MASS $K$-band selected
sample, as assessed by comparison with synthesized $K$-band apparent 
magnitudes of a deeper optically-selected catalog from the SDSS.
The 2MASS K-selected sample is less complete for blue objects.
\label{fig:compl-2mass} 
}
\end{center}
\end{figure}

As described in detail by \citet{mcintosh06}, from a comparison
 to the SDSS main galaxy sample, it turns out that the K-band
 incompleteness of 2MASS is modest 
 and the only major source of incompleteness are blue low surface 
 brightness disk galaxies.
 In order to check the completeness of our particular 2MASS selection 
 we follow a similar philosophy to \citet{bell03} and \citet{mcintosh06},
 by exploring the
2MASS properties of galaxies selected in the optical regime. 
We selected a sample of galaxies with known 
redshifts and SDSS photometry from NED.
 With the same procedure as described in \citet{bell03}, 
the SDSS $u,g,r,i,z$ magnitudes and the redshifts 
were fit with stellar population template spectra
derived using the PEGASE stellar population code \citep[see][for an earlier version of the model]{fioc97}. These template
fits were used to estimate $k$-corrections for the sample, and
permitted estimation of the $K$-band magnitude of a galaxy from the
optical fluxes alone. In this way we could explore whether
2MASS detected all the galaxies which one would {\it a priori} expect
to be detected.
The integrated incompleteness is approximately 5\%.
Fig.~\ref{fig:compl-2mass} shows the color dependence of the completeness 
at $0.01<z<0.03$, showing that the sample is less complete  at bluer colors.

\begin{figure}[t]
\begin{center}
\plotone{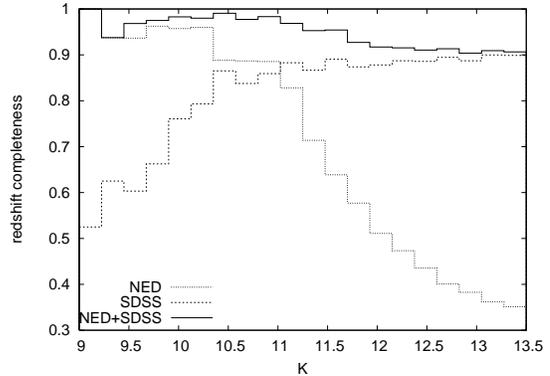}
\caption{K-band magnitude dependence of redshift completeness of the sample. The three histograms show the redshift completeness of the 2MASS-selected sample using data from different sources: The dotted line shows NED completeness only, the dashed line SDSS completeness, and the solid line shows NED and SDSS combined completeness.
\label{fig:compl-z} 
}
\end{center}
\end{figure}

Figure~\ref{fig:compl-z} shows the 
 completeness of the redshift data coming from the two sources.
SDSS is most incomplete at the bright magnitudes, whereas in NED 
the completeness decreases with $K$ magnitude. This trend dominates 
the combined catalog, but the decrease of the completeness is much  
 more moderate.
With our selection criteria, at least
 one of the redshift data (SDSS or NED) is available for  91\% of the
 $K<13.5$ galaxies, the incompleteness mainly
 coming from faint, $K>12$ galaxies. 

For the structural studies we select a nearby subsample which is volume 
 limited between $0.01<z<0.03$. The corresponding and K-band absolute
 magnitude limit
 is $M_K < -22.31$  (which approximately corresponds to
 $L>L^*/5$, \citealp{bell03}). This sample contains 3006 galaxies. Its
 estimated completeness 
 is approximately 90\%; this completeness is dominated by the spectroscopic
incompleteness (largely from SDSS, and this is primarily a geometric effect
owing to the inability to place fibers on all objects) coupled with 
a modest trend to miss out low surface brightness, blue objects; for 
red galaxies, the focus of this work, the sample is essentially limited
only by SDSS fiber placement.

\subsection{Fitting}
\label{sec:fit}

We carried out image fitting with GALFIT to select early-type
candidate systems. 
For fitting the galaxy profiles we constructed an automated pipeline
involving programs SExtractor version 2.3.2 \citep{sextractor} and
GALFIT version 2.0.3. \citep{galfit}. SExtractor creates a catalog of
sources from an astronomical image. The  output parameters can be used to
create an initial setup for light profile fitting with GALFIT. 

We pipeline was run on SDSS $r$-band images.
For each image, SExtractor was used to look up all
sources and to identify the target galaxy, using SDSS positions
  as input. Besides our primary galaxy,
 we fit the nearby 
 objects that could influence the light profile fit of the main
 galaxy. 
The selection of these secondary targets was made by
 distance and intensity: the  object was ignored if the
 distance between the two objects exceeded 1.5 times the sum of 
 their sizes, or the intensity of the secondary object was lower than 1\%
 of the main object. In some cases SExtractor identifies parts of the
 galaxies, such as large H{\sc ii} regions as separate objects. In order
 to correct these cases, check images were created with labeled fit targets. The images were visually inspected and unnecessary
 sources were untargeted.
These results were used to create the
 GALFIT initial setup. 

The main target galaxies were fitted  with a 
\citet{sersic} profile 
\begin{equation}
\Sigma_b(r)=\Sigma_e e^{-c(r/r_e)^{1/n} -1}
\label{eq:sersic}
\end{equation}
where $r_e$ is the effective radius of the galaxy, $\Sigma_e$ is the
surface brightness at this radius, $n$ is the S{\'e}rsic index
(in the
special case when $n=1$ we get the exponential profile, $n=4$ corresponds
to de the Vaucouleurs profile)
and $c=1.9992 n^{-0.3271}$ so that the half of the flux is within $r_e$.
 The nearby objects (secondary targets) were fitted
either by S{\'e}rsic or, if more star-like, by the point spread function (PSF). 
The PSF was created from stars of each
 field image separately, based on Moffat profile fits in the form:
\begin{equation}
\Sigma_b(r)=\frac{\Sigma_0}{[1+(r/r_d)^2]^m},
\label{eq:moffat}
\end{equation}
where $r_d$ is the dispersion radius and $m$ is the power-law index.

Initial values for sky level, flux, size and axis ratio were set 
 using SExtractor output parameters. The S{\'e}rsic index and axis ratio
 for galaxies
were set to initial values common for all
 objects, $n=  1.5$, $b/a=0.9$. 
 The fit parameters were constrained using empirical
 (observational) and  SExtractor output information:
$n$ was required to lie in the interval $0.5<n<7$, 
axis ratio $0.3<b/a<1$, position within 2 pixels
of the SExtractor value, and half-light radius 
between 0 and 1.5$r_{\rm SExtractor}$.
The
   pipeline first fits the objects separately, keeping the other
   objects fixed, and then finally fits the objects all together in the end, 
using their separate fit results as the initial guess. The sky is fixed
   at all times.

One way to test the fitting pipeline is to fit simulated objects of
known parameters. We used a suite of simulated galaxies composed
 from a S{\'e}rsic bulge + 
exponential disk, with a wide range of magnitudes, bulge to total
ratios and light profile parameters. We selected a test sample similar 
to our sample of interest: bulge dominated systems, having bulge to total 
ratios $B/T>0.6$ and magnitudes brighter than $16.5$ (the $K<13.5$ cut 
corresponds approximately to $r<16.4$). 
The simulated galaxies were fit by the pipeline described above.
The fits give us an important sanity check of the fitting
procedure.  Our results show that S{\'e}rsic fit works
relatively well,
the simulated magnitudes are reproduced with an 
error $\sim 0.2$\,mag for the test sample. 

\begin{figure}[t]
\begin{center}
\plotone{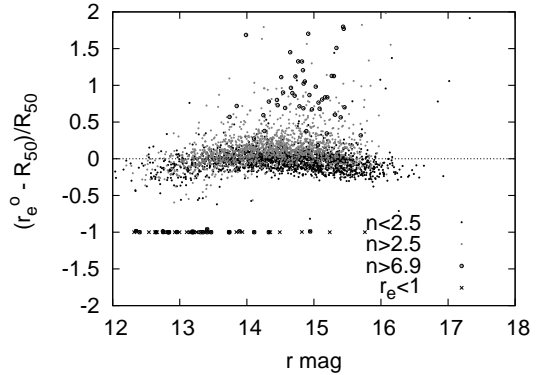}
\caption{The relative difference of the fitted  circularized S{\'e}rsic
 $r_e$  vs. SDSS  $R_{50}$ as a function of $r$ magnitude. Small
black symbols show $n<2.5$ galaxies (a proxy for disk-dominated galaxies; $n=1.5$ excluded), while small gray symbols show systems with
$n>2.5$ (a proxy for bulge-dominated galaxies). Larger
 open circles show systems where the fit hit the upper limit in $n$, 
 and crosses show where the fit hit the lower limit in $r_e$.
\label{fig:fit-sersic-sdss} 
}
\end{center}
\end{figure}

When fitting the real images, the S{\'e}rsic fit is
successful and gives results similar to the SDSS photometric
parameters for the majority of our galaxies. 
 It also corrects the known error of the SDSS imaging pipeline that it
 tends to split
up large objects into multiple sources. This affects 4.6\% of the
objects in our sample, mostly objects with $n<1$. The
S{\'e}rsic fit fails in 1.5\% of the cases because  
the fit did not converge or the image file was corrupt. 1.9\% of the
fit results are unreliable because the fit did not move out from the
initial setup or reached the fitting boundary. 
Figure \ref{fig:fit-sersic-sdss} shows the fitted sizes (circularized
$r_e^o = r_e \sqrt{b/a}$) compared with the SDSS $R_{50}$ radii. 
The
fit gives results consistent with the SDSS sizes within 15\%, when 
$>3\sigma$ outliers are discarded. 

We choose $n>2.5$ galaxies for further study, 
which appears to be a reasonable way to automatically 
reject a significant fraction of the disk-dominated galaxies
\citep{bell_gems}\footnote{If instead a selection by $r$-band concentration as calculated by the SDSS (defined as the ratio of the aperture containing 90\% of the flux in the Petrosian radius to the half-light radius) being larger than 2.6 \citep{strateva01} was used, the sample would remain unchanged at the 90\% level.}.  We further exclude all objects with failed
fits, as described above.  The criterion $n>2.5$ is deliberately 
generous, and includes a large number of galaxies with prominent
disks \citep{bell_gems}.  Accordingly, we classified 
the $n>2.5$ subsample by eye, classifying
553 objects as elliptical (pure bulge; E), 498 as lenticular (smooth, non-star-forming disk; S0), 479 as later types (Sa-Irr)
 and 27 as mergers. 

   We choose to visually exclude all galaxies with stars or
     companion/projected galaxies which made a major contribution to the
     asymmetry in the area in which asymmetry is measured.  The spirit of
     this cut is to excise all systems where we were concerned that the
     asymmetry measurement would be corrupted by a non-interacting (or
     non-strongly-interacting) object (be it a projection or an
     interaction).  While some form of masking might have been desirable,
     we adopted this conservative approach to account for objects missed
     by SExtractor and because the extent of the masked area is never
     straightforward to choose.  We lose a little over 
     1/3 of the sample using this cut
     (it is our most important cut), and the loss is primarily geometric
     (and therefore introduces no bias) for the galaxies lost from bright
     stars and projections, and those galaxies thrown out because of a
     pre-merger satellite.  We have confirmed that the color--magnitude diagram of these excised galaxies is very similar to the final sample of galaxies without projections (there are a few more outliers in the redwards and bluewards directions in the excised sample compared to the final sample, as is expected from a sample of galaxies where there is a projected star or galaxy contaminating the photometry). 
There is some natural fuzziness in this
     process, and some advanced mergers where the second nuclei (and large
     amounts of tidal disturbance) were within the asymmetry measurement
     area were included in the sample.
After this final selection our sample contains 288 E and 319 S0 type
galaxies, together 607 objects.

\section{Asymmetry and stellar population measurements}
\label{sec:color-asymm}

 For the investigation of the correlation between the post-merger structure and 
 stellar 
population we need to construct two broad classes of metric: descriptions
of the (hopefully tidally-induced) non-equilibrium structure of an 
early-type galaxy (\S \ref{sec:asymm}), and parameters characterizing 
the stellar population, like color, age, and metallicity (\S \ref{sec:color-mag}).

\subsection{Structure}
\label{sec:asymm}

   We choose to use asymmetry as a repeatable diagnostic of 
non-equilibrium structure. This measure is insensitive to symmetric 
bars and rings (unlike the residual structure measures used by 
\citealp{tal09}), 
which are important structures in some S0s but are not (at least unique) 
indicators 
of a recent perturbation. 
 Our metric is not as tailored as the fine structure parameter used by
 \citet{ss92}, but the asymmetry measurement is
repeatable and can be automatically applied to survey datasets.

\begin{figure}[t]
\begin{center}
\plotone{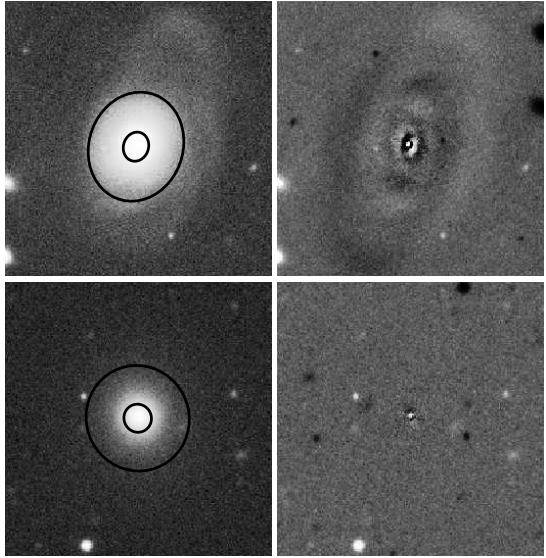}
\caption{Two examples of asymmetry measurement. Left panels: the original image,
  with the elliptical annuli over-plotted; right panels: asymmetry images.
\label{fig:asymm-def} 
}
\end{center}
\end{figure}

We adopt an asymmetry measure similar to methods discussed by 
\citet{conselice00}. 
We measure asymmetry within an elliptical annulus of 0.8--3
$r_e$. In this way we minimize the impact of the bright galaxy
core on our asymmetry measure which would otherwise be
 very sensitive to centering uncertainties and the influence of dust disks and features in the inner part of early-type galaxies\footnote{Note that initially we chose to adjust the center to minimize asymmetry within 0.5$r_e$.  Galaxies with significant asymmetry often showed that the center of the brightest isophotes was offset from that of the fainter isophotes, yielding a change in the asymmetry with radius.  We found that the GALFIT center was very close to a choice that minimized $A$, with good S/N.  Accordingly, in what follows, we choose to use the GALFIT center, and omit the central regions from the asymmetry measure.  }.  
 A different choice of the aperture
 does not influence the results qualitatively, unless the inner
 $0.5 r_e$ is included in the asymmetry measure.
The asymmetry calculating algorithm is carried out as follows
 in the steps 1-7, for an illustration see Figure \ref{fig:asymm-def}.
\begin{enumerate}
\item Rotate the image around the GALFIT center by 180$^\circ$. 
\item Subtract the rotated image from the original one.
\item Integrate flux in quadrature within the elliptical annulus between
0.8-3 $r_e$.  This measure will have an additive 
contribution from noise which can be
separated and corrected for. Let us separate the flux $f$ into signal
and noise term, $f=s+n$ and analogously for the rotated image
$f'=s'+n'$. Then the asymmetry flux $f-f'$ 
integrated in quadrature over the image will read 
\begin{eqnarray}
\sum(f - f')^2 & =& \sum(s +n - s' - n')^2 \nonumber\\ 
&=&\sum(s-s')^2 + \sum(n-n')^2 \nonumber\\
&=&\sum(s-s')^2 + \sum n^2 + \sum n'^2,
\end{eqnarray}
as signal and noise, as well as the both noise terms are 
uncorrelated. This will result in a non-zero
asymmetry even in the case of $s=s'$, i.e. perfectly symmetric
underlying signal. We can correct for  
this by estimating the noise contribution and subtracting it
from the integral. We estimate $n-n'$ using the GALFIT model
image with noise similar to the observed data. 
\item Apply Poisson noise + read noise to the (symmetric) GALFIT model image.
\item  Measure asymmetry in the same way as for the actual image (steps
1-3, rotate, subtract, integrate in quadrature).
\item Subtract model asymmetry from the object asymmetry.
\item In the end we make the asymmetry measure independent of the total
flux: normalize by the object (sky subtracted image) flux squared and
integrated within the same area. 
\end{enumerate}

To summarize, our asymmetry measure  is defined as 
\begin{equation}
A= \left(\sum(f - f')^2  -  \sum(m-m')^2\right) / \sum(f^2).
\end{equation}
where $f$ denotes the flux of the galaxy, $m$ is the fitted S{\'e}rsic
model, the primes denote the corresponding quantities rotated by 180
degrees, and the summation goes over pixels within the elliptical
annulus 0.8-3 $r_e$.  Because the asymmetry is determined from 
the square of the asymmetry image, values of $\sim$ 0.01 are 
already very highly significant, suggesting asymmetries involving $\sim 5$\% 
of the flux (this definition counts asymmetric flux from both the positive
and negative parts of the asymmetry image; other definitions 
divide the asymmetry by two to account for this effect).

The uncertainties in asymmetry were calculated in two complementary ways.
Random uncertainties were estimated by subjecting symmetric model galaxies 
with realistic model and sky noise properties to the same procedure above; these random uncertainties are $\delta A \sim 0.0003$, and are shown by the 
horizontal component of the cross-shaped error bars in Figs.\ \ref{fig:asymm-color}, \ref{fig:asymm-age} and \ref{fig:asymm-met}.  The second estimate of asymmetry uncertainty is systematic in nature, reflecting our choice to neglect the inner 0.8 $r_e$ when estimating the asymmetry (in order to minimize the effect of small centering uncertainties, and the influence of small dust lanes or disks in the inner parts of these early-type galaxies).  We changed the inner cutoff to 0.6-1 $r_e$ and the outer cutoff to 2-4 $r_e$; the result was that 
the asymmetries changed significantly, because of the different area being 
probed, but to first order all galaxies 
moved to higher or lower values of asymmetry together.  Experimentation 
revealed that the best way to visualize this source of uncertainty is 
as an uncertainty in asymmetry rank.  The rank changed typically by $\sim 5$\%; we assign an asymmetry rank error of 5\%, and for the purposes of visualization we translate this rank uncertainty into an asymmetry uncertainty using the asymmetries of galaxies that have ranks 5\% different from the galaxy of interest.  We show error bars calculated in this way in Figs.\  \ref{fig:asymm-color},  \ref{fig:asymm-age} and \ref{fig:asymm-met} as the solitary horizontal error bars.  The $A$ uncertainties from this source are a strong function of $A$, increasing from very small values for low values of $A$ to large values for the tail of 
galaxies extending to large $A$.  

\subsection{Stellar population diagnostics}
\label{sec:color-mag}

We choose for consideration in this work two types of stellar
population diagnostics: galaxy colors, as an indicator
of broad changes in the stellar population; and 
luminosity-weighted stellar ages and metallicities --- available
for roughly 2/3 of the sample --- as model-dependent but more 
information-rich stellar population diagnostics.

 We use SDSS model $g-r$ color corrected for foreground extinction and AB offset, then k-corrected using method of \citet{bell03}.  Note that these
colors are very similar to colors we measured ourselves
within $0<r/r_e<3$, and the results change only imperceptibly when such 
colors are used.  We adopt the model colors because these are readily and
publicly available.  We assign uncertainties of 0.03 mag in g-r color to account for random photometric errors (sub-dominant) and calibration uncertainties from galaxy-to-galaxy in the zero-points of the SDSS photometry (the dominant contribution). 
The color combination $g-r$ was used because it is the bluest 
well-measured SDSS color
combination.  While, in principle, $u-r$ color is more sensitive to stellar
population (as it straddles the 4000{\AA} break), $u$-band 
has poor S/N, and suffers from increased systematics compared
to $g$-band.

In order to aid in disentangling the effects of the age/star formation
history of a galaxy from metallicity in driving its color, we use
estimates of luminosity-weighted age and metallicity from \citet{gallazzi05}. 
\citet{gallazzi05} derived ages and metallicities from all
SDSS spectra with median S/N$>20$ per pixel.  This stringent S/N cut
selects only $\sim 1/4$ of all galaxies, primarily those with high surface
brightness cores.  Our sample consists of E/S0 galaxies only, and
therefore is much more complete than the whole sample: 97\% of the Es 
and S0s in the sample with SDSS spectra in DR4 have age and metallicity 
estimates from \citet{gallazzi05}.

Ages and metallicities are derived by comparing observed spectral
absorption features (three primarily age-sensitive indices, 4000{\AA}
break strength D4000, H$\beta$, H$\gamma_A +$ H$\delta_A$, and two
primarily metallicity-sensitive indices $[{\rm Mg_2 Fe}]$ and $[{\rm
MgFe}]'$) with a library of \citet{bc03} models with a wide
range of star formation histories and metallicities.  The models have 
solar abundance ratio patterns (i.e., they do not allow for 
$\alpha$-element enriched abundance patterns), but the impact of this 
shortcoming on the derived ages and metallicities is not that severe owing 
to the choice of combined Magnesium and Iron indices as the 
metallicity-sensitive indices (see \citealp{gallazzi05} for details and 
more discussion).  In this paper, the main emphasis is on the age 
estimates (recall that the hypothesis being tested is that 
richly-structured early-type galaxies have younger ages), which depend 
primarily on the Balmer lines and the strength of the 4000{\AA}-break.  
These ages are {\it luminosity-weighted}, i.e., the ages are the mean 
luminosity-weighted ages of the models that provided a reasonable fit to 
the five observed spectral indices, and are correspondingly younger than 
the mass-weighted ages of those model populations.  

Section 2.4 of \citet{gallazzi05} discusses in detail the sources of 
random and systematic error in these age and metallicity estimates.
Briefly, random errors (from the spectra themselves) do not cause
any systematic shifts in the median ages or metallicities from their
correct values (verified using simulations), and for our subsample
lead to a typical uncertainty in log age of 0.1 dex and metallicity of 
0.09 dex.  The main systematic errors were identified to be 
alpha element overabundance uncertainties and the degree of 
burstiness in the star formation history of the galaxy population.  
The systematic errors in metallicity and log age from these
sources was argued to be less than 0.05 dex for relatively large 
variations in alpha element overabundance and the fraction of galaxies
undergoing large bursts (we will discuss further these uncertainties
in \S \ref{agemet}). 
It is worth noting 
that the luminosity-weighted ages are reasonably model-dependent, and depend
on when star formation is assumed to start and to the timing 
and amplitudes of bursts of star formation.  For example, a model with 
given line strengths could be fit by a population that started forming stars 
12\,Gyr ago, or only 5\,Gyr ago; this will naturally lead to dramatic
differences in luminosity-weighted age.  Accordingly, in this work, we
use luminosity-weighted ages primarily 
as a tool to understand {\it relative} trends in 
age, and to attempt to disentangle the effects of age and metallicity
on broad-band colors.

\section{Results}
\label{sec:results}

\begin{figure}[t]
\begin{center}
\plotone{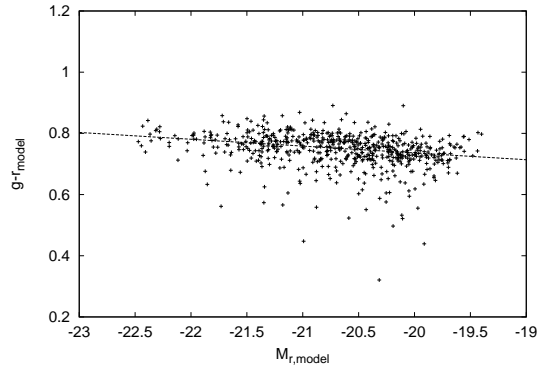}
\caption{The CMR using $g$ and $r$-band SDSS model 
magnitudes for galaxies in our sample with measured asymmetries and 
meaningful SDSS model magnitudes.
The line shows the best fit to these data. 
\label{fig:color-mag} 
}
\end{center}
\end{figure}

\begin{figure*}[t]
\begin{center}
\includegraphics[width=14.0cm]{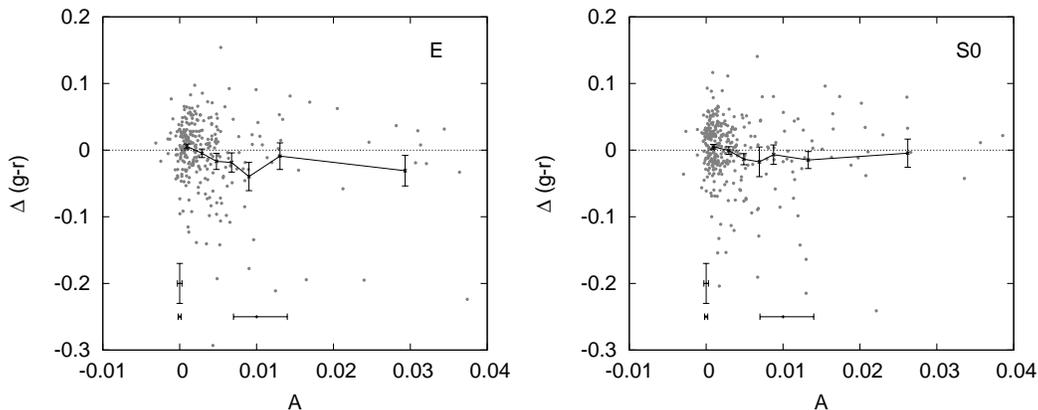}
\caption{Color-asymmetry diagram. Left panel: Es in our sample, right
  panel: S0s in our sample. The lines show averages in bins, with error bars denoting the error of the average. 
The cross-shape error bars indicate the random error 
of asymmetry at A=0 
(see Sec.~\ref{sec:asymm}) 
and the color error. The systematic uncertainty due to the choice of  
 the asymmetry measurement aperture is shown as solitary horizontal error 
bars at two different asymmetry values.
A correlation between color and asymmetry
 is present in both galaxy types.
\label{fig:asymm-color} 
}
\end{center}
\end{figure*}

Figure~\ref{fig:color-mag} shows the CMR for the
605 E/S0 galaxies in our sample with measured asymmetries
and SDSS Model magnitudes.
 We fitted a line to the main locus of the 
CMR,  excluding blue outliers at $g-r<0.6$ from the fit.
The equation of the fitted line is
\begin{equation}
 g-r = -0.023\, M_r + 0.28.
\end{equation}
We subtract this linear relation from the color to get the color offset 
from the red sequence: $\Delta (g-r)$.
The scatter of  $\Delta (g-r)$ is $0.041$ with $3 \sigma$-clipping 
applied; much of this scatter
is intrinsic (not measurement error, the typical measurement error
in model $g-r$ for our sample is {\bf $\sim 0.03$}\,mag and is dominated by 
flat fielding and calibration uncertainties; see also
\citealp{ss92} and \citealp{ruhland07} who also measure similar 
intrinsic scatter in the CMR\footnote{for the purposes of comparing 
scatter in $g-r$ with $U-V$, note that $\Delta (U-V) \sim 2.5 \Delta(g-r)$}).  
The mean color offset is slightly negative, 
$\langle \Delta (g-r) \rangle = -0.005$,  because of the 
clipping of very blue outliers from the CMR fit.

\begin{table*}
\caption{Correlation Probabilities of stellar population--asymmetry
correlations\label{tab:sig}}
\begin{center}
\begin{tabular}{l l l l l l}         
\hline
\hline
Sample & $N$ &  $r$ &  $P(t<t_{\rm obs})$ & $s$ & $P(s<s_{\rm obs})$ \\
\hline
\multicolumn{6}{c}{Color--asymmetry} \\
\hline

E   & 288 & -0.203 & 0.00026 & -0.114 & 0.027 \\
S0  & 319 & -0.0851 & 0.065 & -0.111 & 0.024 \\
all & 607 & -0.161 & 3.5$\times 10^{-5}$  & -0.111 & 0.0031 \\ 
\hline
\multicolumn{6}{c}{Color--asymmetry $|A|<0.01$ and $|\Delta(g-r)|<0.15$ } \\
\hline
E   & 250 & -0.169 & 0.0037 & -0.107 & 0.046 \\
S0  & 275 & -0.139 & 0.010 & -0.121 & 0.023 \\
all & 525 & -0.157 & 0.00015 & -0.114 & 0.0045 \\ 
\hline
\multicolumn{6}{c}{Age--asymmetry} \\
\hline
E   & 190 & -0.37 & 7.4$\times 10^{-8}$ & -0.363 & 2.9$\times 10^{-7}$ \\
S0  & 199 & -0.316 & 2.8$\times 10^{-6}$ & -0.375 & 6.4$\times 10^{-8}$ \\ 
all & 389 & -0.351 & 5.4$\times 10^{-13}$ & -0.376 & 6.2$\times 10^{-14}$ \\
\hline
\multicolumn{6}{c}{Metallicity--asymmetry} \\
\hline
E   &190 & -0.22 & 0.0012 & -0.0486 & 0.25 \\
S0  &199 & -0.073 & 0.15 & -0.0178 & 0.40 \\
all &389 & -0.173 & 0.00031 & -0.0235 & 0.32 \\ 
\hline
\end{tabular}
\\
\tablecomments{Probability of observing a given stellar population--asymmetry
correlation 
in a completely uncorrelated dataset, as evaluated using 
Pearson's correlation coefficient $r$ and Spearman's rank correlation 
coefficient ($s$).  The probability of a given 
value of the correlation coefficient $r$ being due to chance alone 
in an uncorrelated dataset is distributed as a Student's $t$ distribution with 
$N-2$ degrees of freedom, where
$t = r/\sqrt{(1-r^2)/(N-2)}$.  
} 
\end{center}
\end{table*}

\begin{figure*}[t]
\begin{center}
\includegraphics[width=14.0cm]{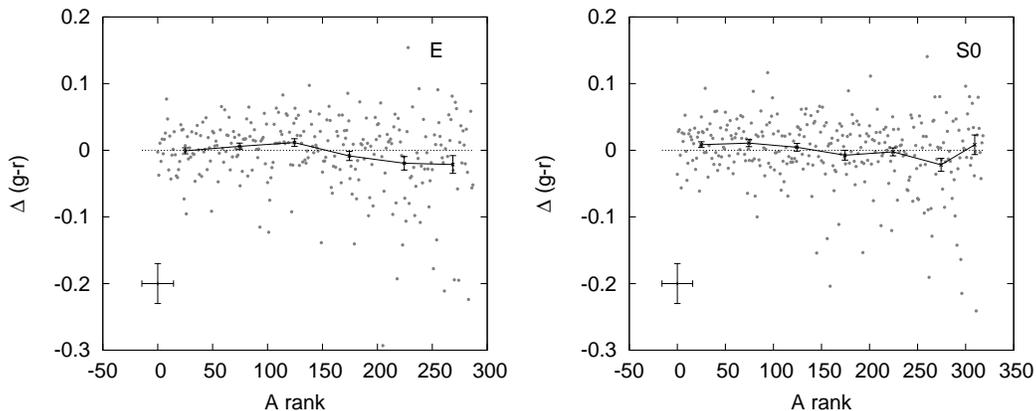}
\caption{Color offset $\Delta (g-r)$ shown as a function of the rank
of asymmetry. Left panel: Es, right
  panel: S0s. Symbols are same as in Fig.~\ref{fig:asymm-color}, 
the cross shaped error bars show the typical systematic uncertainty of 
asymmetry (see explanation in Sec.~\ref{sec:asymm}) and the color error.
\label{fig:asymm-color-sort} 
}
\end{center}
\end{figure*}

\begin{figure}[t]
\begin{center}
\includegraphics[width=8cm]{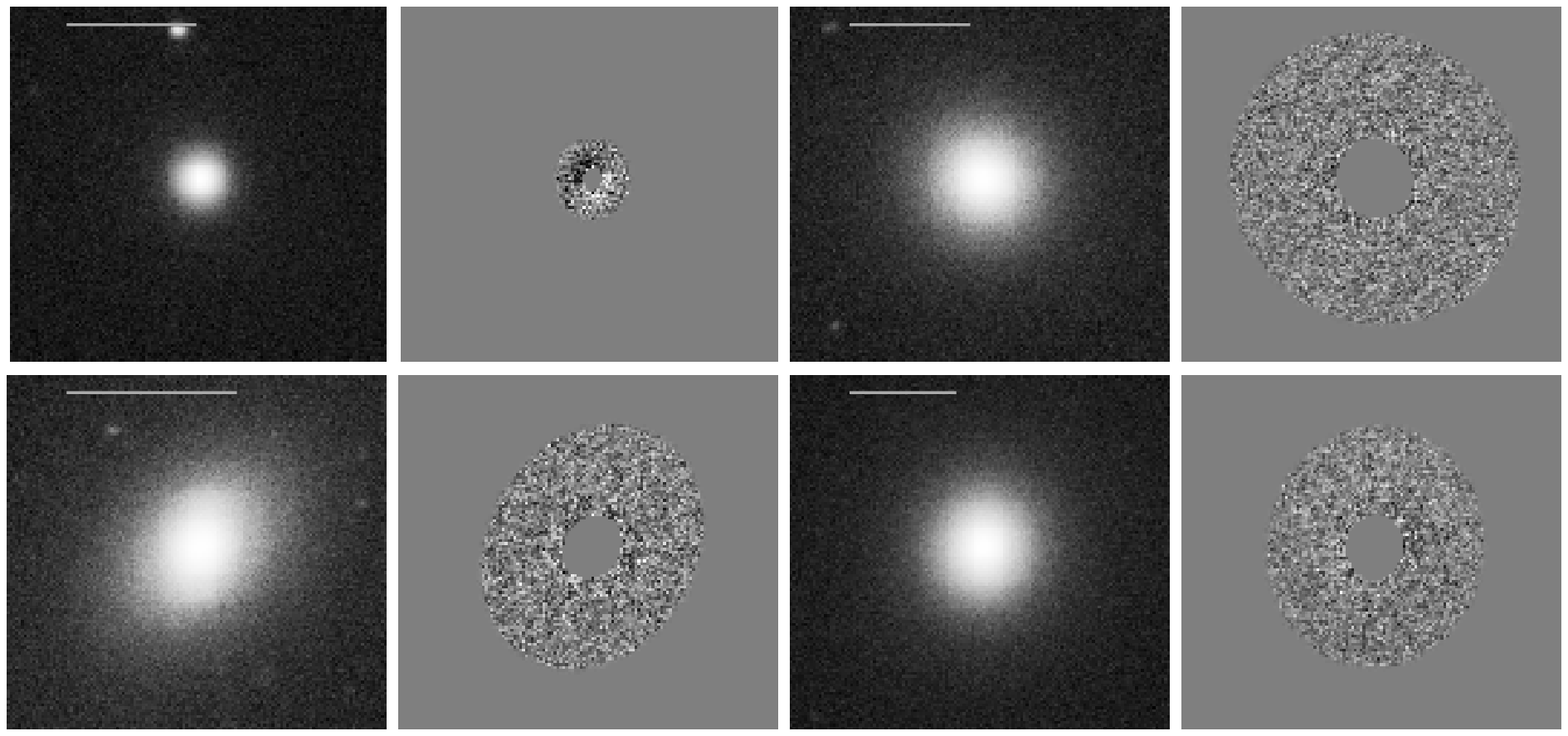}
\includegraphics[width=8cm]{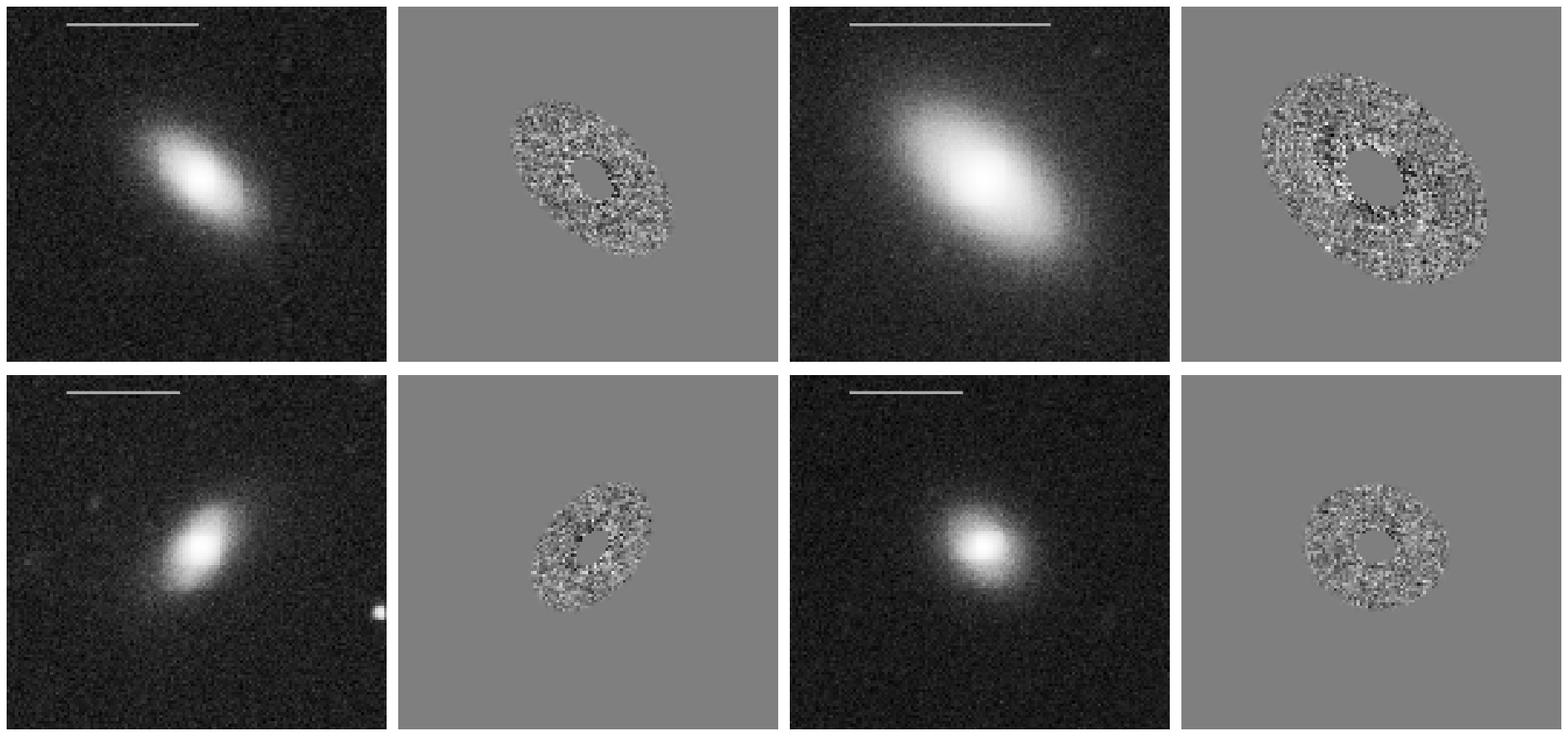}
\caption{Galaxy images (first and third column) with their corresponding asymmetry images in elliptical annuli (second and fourth column, respectively) for the $0<A<0.001$ asymmetry bin (most symmetric objects).  First and second row: Es, third and fourth row: S0s. Each box is approximately 55 arcsec on a side. A 10 kpc horizontal line is shown for each object. 
\label{fig:asymm-low} 
}
\end{center}
\end{figure}

\begin{figure}[t]
\begin{center}
\includegraphics[width=8cm]{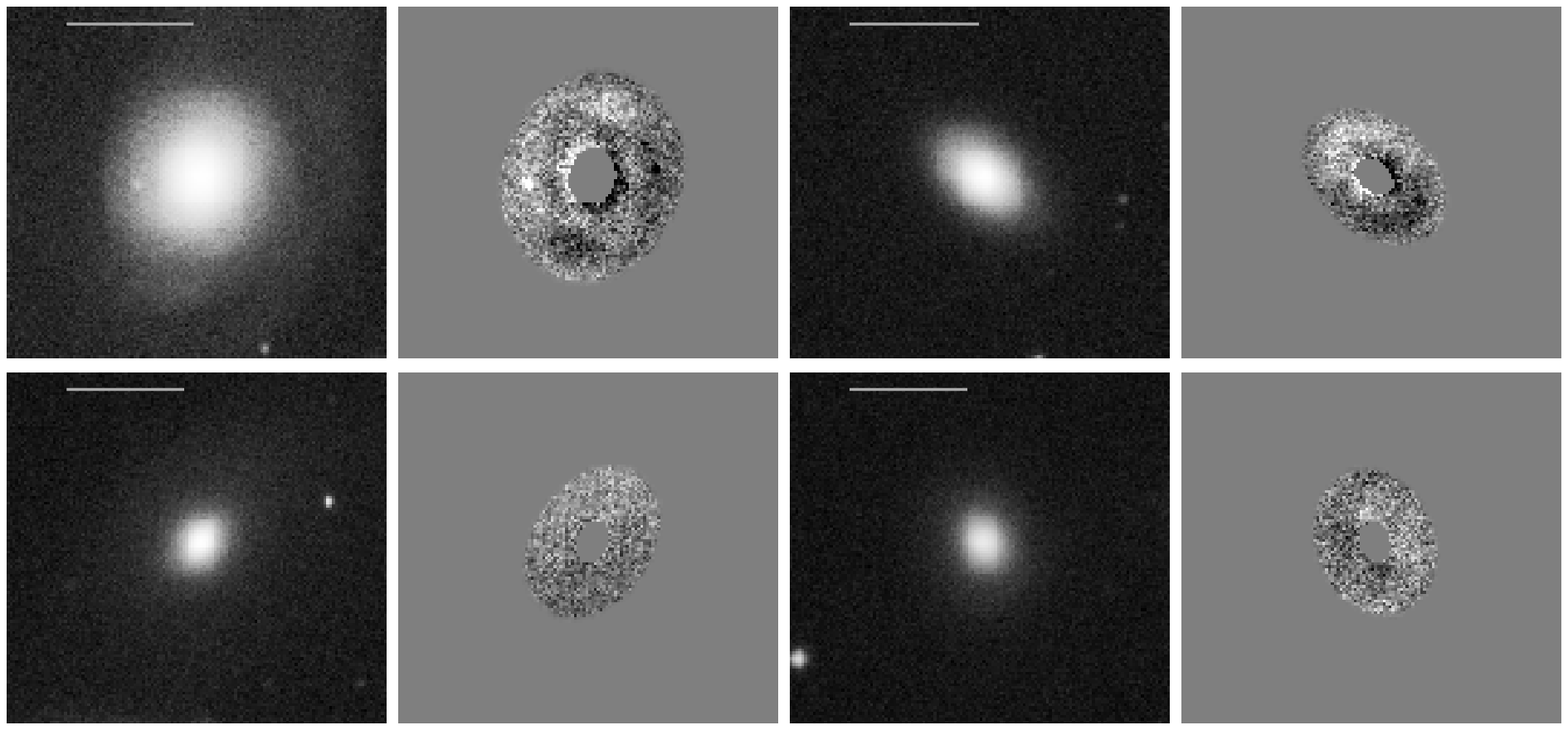}
\includegraphics[width=8cm]{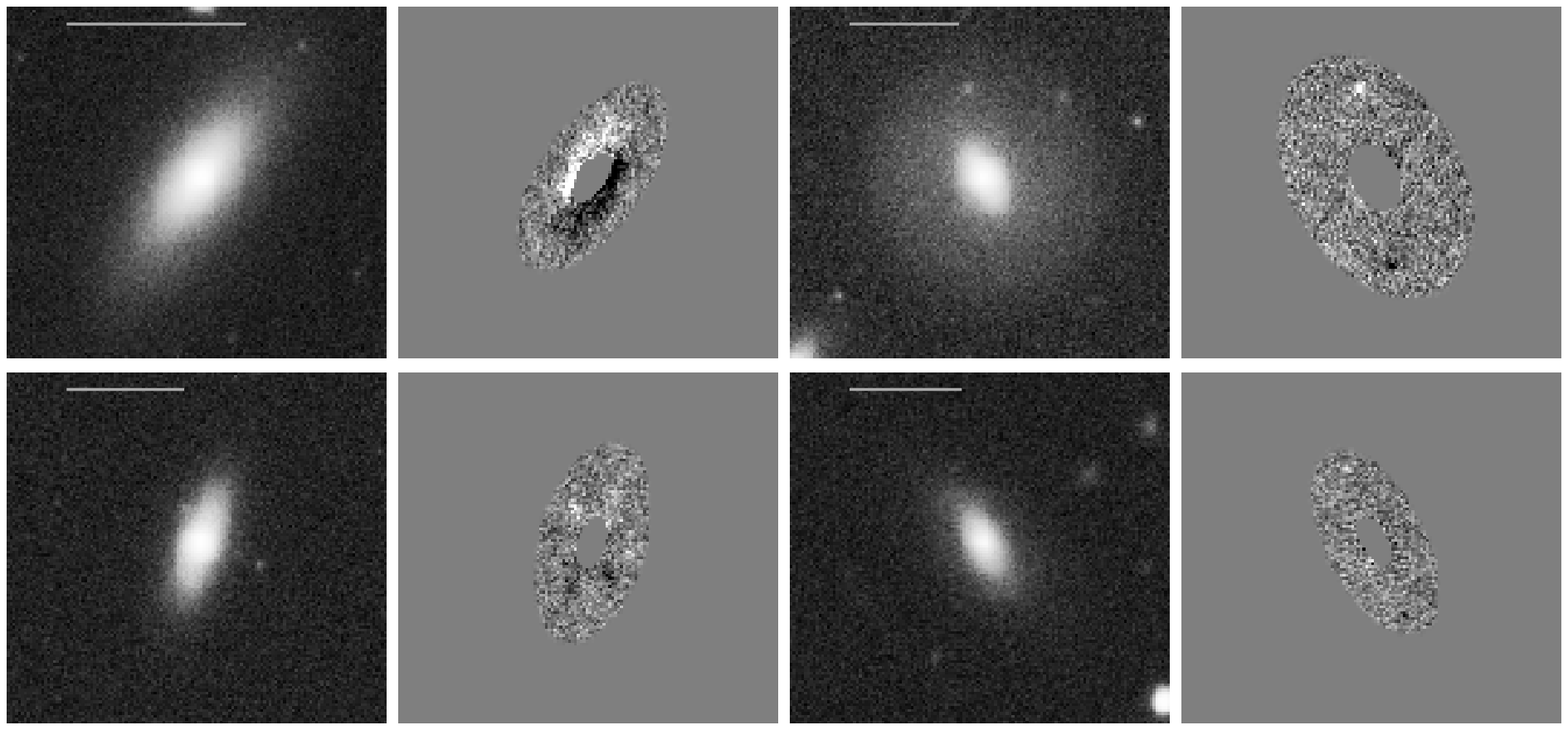}
\caption{Same as Fig.~\ref{fig:asymm-low} but asymmetry range $0.005<A<0.01$. 
These are typical asymmetry values for the galaxies that are
driving the color--asymmetry correlation.
\label{fig:asymm-main} 
}
\end{center}
\end{figure}

\begin{figure}[t]
\begin{center}
\includegraphics[width=8cm]{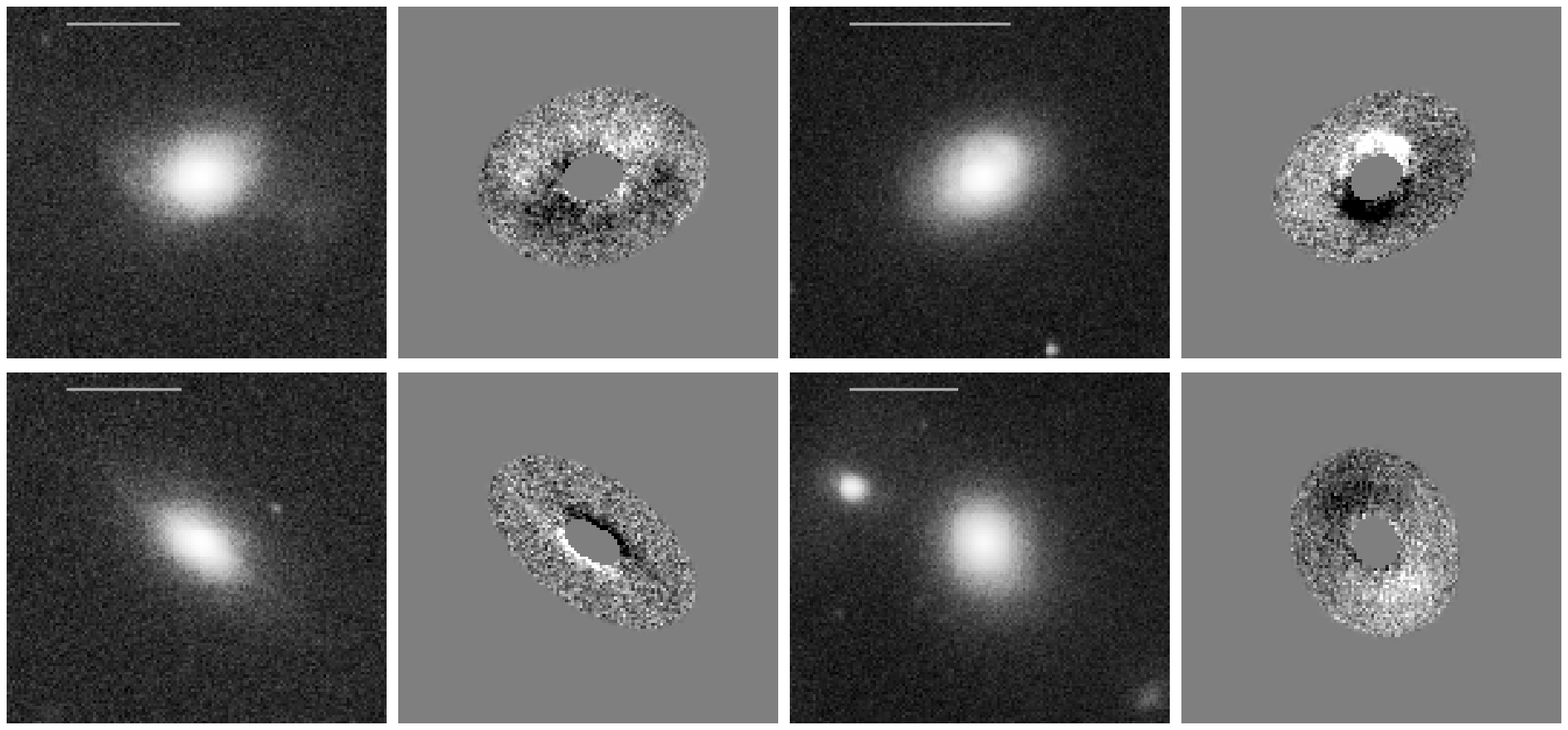}
\includegraphics[width=8cm]{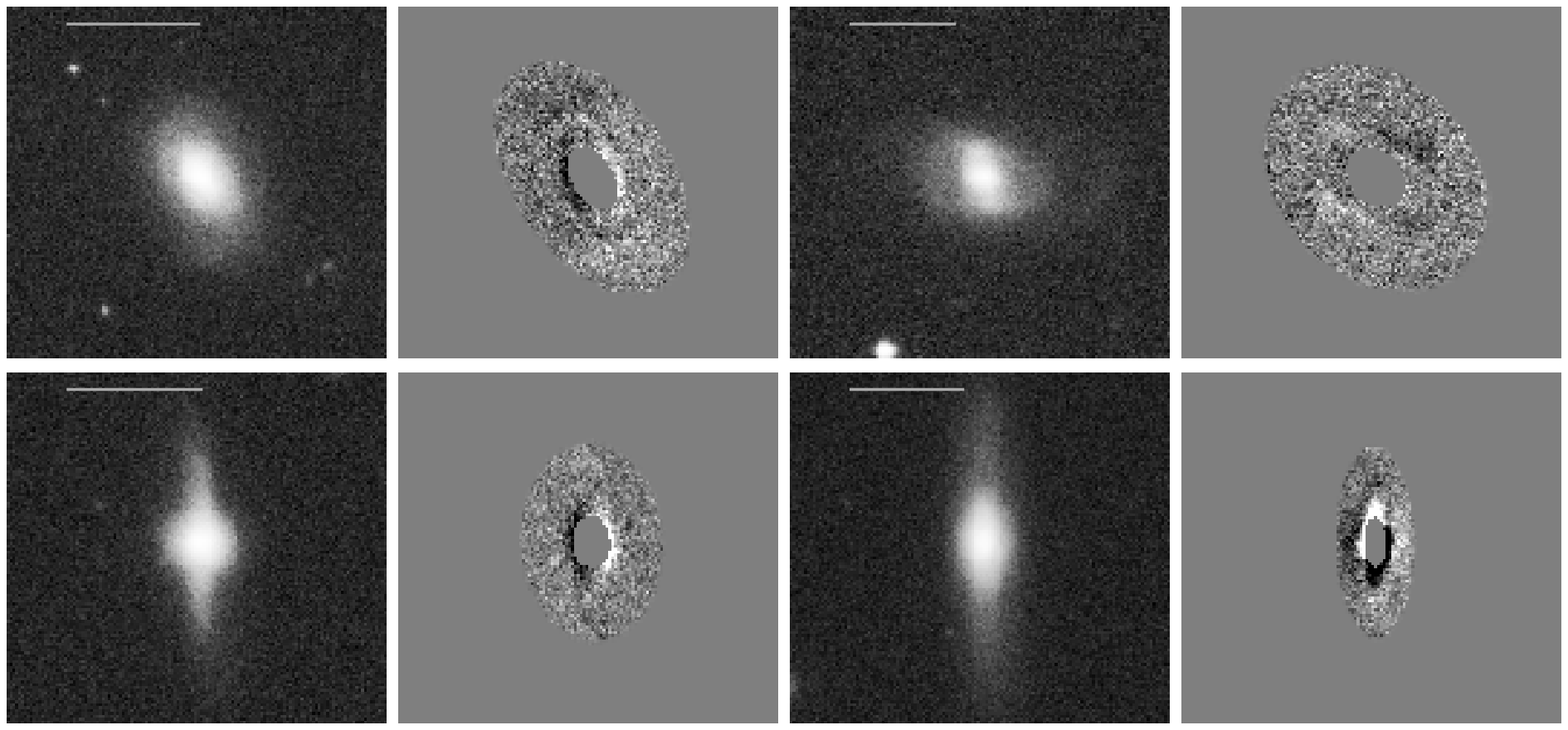}
\caption{Same as Fig.~\ref{fig:asymm-low} but asymmetry range $0.02<A<0.04$ -- most asymmetric E/S0 galaxies.
\label{fig:asymm-high} 
}
\end{center}
\end{figure}

Figure \ref{fig:asymm-color} shows the color offset $\Delta (g-r)$ 
as a function of our asymmetry measure, 
with bin averages plotted with lines
(5 bins equally spaced between $A$ of 0 and 0.01, 
a bin with $0.01<A\le 0.02$ and one with $0.02<A\le 0.04$). 
The data show a correlation between 
asymmetry and color offset for both E and 
S0 galaxies, with approximately the same 
slope and zero point; this figure can be compared 
reasonably directly with Figs.\ 2 and 3 of \citet{ss92}, 
and confirm their detection of a significant correlation 
between substructure in early-type galaxies and the offset from 
the CMR.

One obvious feature of this correlation is that it is very
scattered; the scatter in the CMR exceeds the magnitude of the correlation
between color offset and asymmetry.  This scatter
is primarily intrinsic (i.e., not measurement error; see for example
\citealp{gallazzi06} for a demonstration that the stellar populations
of galaxies on the red side of the CMR differ substantially from those
on the blue side).  We explore this issue in 
sections \ref{agemet} and \ref{interp} in some depth.

The correlation is clearest for $A<0.01$ --- the run of both the
mean and median color at $A<0.01$ is very similar.   At larger asymmetry 
values, in particular for elliptical galaxies, there are quite a 
few galaxies with significant asymmetry but relatively red colors --- the 
median color for $A>0.02$ ellipticals is reasonably red.  We discuss
this later in \S \ref{sec:dry} in the context of dissipationless
merging between already gas-poor early-type galaxies.
Nonetheless, the bulk of the early-type galaxy population (those with $A<0.01$)
shows a trend between color offset and asymmetry. 
To help visualize the trend, we ranked the E/S0 galaxies
by $A$ and show how rank $A$ correlates with color 
in Fig.~\ref{fig:asymm-color-sort}.  
We have investigated the significance of the correlation between 
$\Delta (g-r)$ and $A$ using Pearson's correlation coefficient $r$
and Spearman's rank correlation coefficient $s$ (see
Table \ref{tab:sig} for the full breakdown by galaxy type
and stellar population diagnostic)\footnote{
We present one-sided probabilities as we test the significance
of the observed (and expected) anticorrelation between 
stellar population diagnostics (color/age/metallicity) and asymmetry.}.
The correlation between
$\Delta (g-r)$ and $A$ exists but has large scatter.
Over the full range of color and asymmetry, the correlation 
between $\Delta (g-r)$ and $A$ has less than a 0.0035\% probability of
being from chance alone as estimated using Pearson's correlation 
analysis (or a 0.015\% probability of being 
from chance alone if one restricts attention to relatively
low asymmetries and small color offsets).  If instead one studies 
ranked color offsets and ranked asymmetries, a Spearman's rank correlation
analysis argues that the relationship between asymmetry and color for 
the combined E/S0 sample
has less than 0.5\% probability of being given by chance alone\footnote{This 
is a relatively modest significance, and if this were the only
set of stellar population diagnostics, one may be concerned about 
the existence of this correlation.  We will illuminate this 
issue later in \S \ref{agemet}.}.  In all cases, the values of 
correlation coefficient and significance change little if different 
contributions of inner/outer radii are used for defining the asymmetry values. 

In Figures 
\ref{fig:asymm-low}, \ref{fig:asymm-main}, and 
\ref{fig:asymm-high} we show several examples of different asymmetry values for both morphological types, to allow the reader to generate some 
intuition for the asymmetries shown by the sample. 
In Fig.~\ref{fig:asymm-low} shows a sample most symmetric 
Es and S0s, with asymmetry in the range $0<A<0.001$.
Fig.~\ref{fig:asymm-main} shows galaxies with  $0.005<A<0.01$; 
galaxies in this range are those that drive the bulk of the 
correlation between color offset and asymmetry.
Fig.~\ref{fig:asymm-high} shows highly asymmetric galaxies from the asymmetry bin $0.02<A<0.04$.
It is clear that the asymmetries being discussed here are not 
from isolated H{\sc ii} regions, or low-level spiral arms; rather, 
these asymmetries are to a great extent driven by 
large-scale asymmetries, many of them matching qualitatively the form
expected for tidal features.   Taken together
with Fig.\ \ref{fig:asymm-color}, these figures underline the primary
result of this work (and \citealp{ss92}; \citealp{tal09}): early-type galaxies
with asymmetric signatures of tidal interactions tend to be offset
from the locus of the CMR, in the sense that 
more asymmetric galaxies tend to be bluer than undisturbed galaxies
(albeit with large scatter around this trend).

\subsection{What drives these correlations, age or metallicity?}
\label{agemet}

\begin{figure*}
\begin{center}
\includegraphics[width=14.0cm]{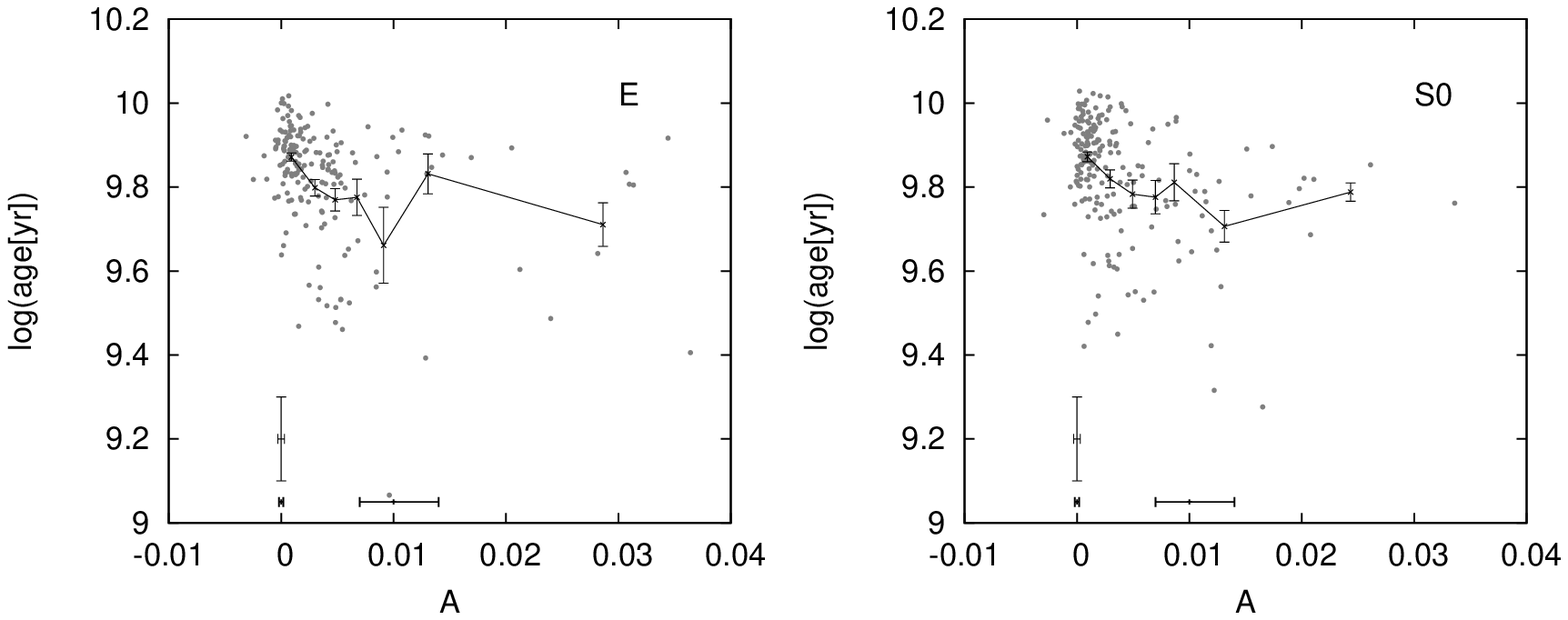}
\caption{The logarithm of the luminosity weighted age (in years) versus asymmetry.
Left panel: elliptical galaxies, right  panel: lenticular (S0) galaxies. 
Symbols, lines and error bars are as in Fig.~\ref{fig:asymm-color}.
\label{fig:asymm-age} 
}
\end{center}
\end{figure*}

\begin{figure*}
\begin{center}
\includegraphics[width=14.0cm]{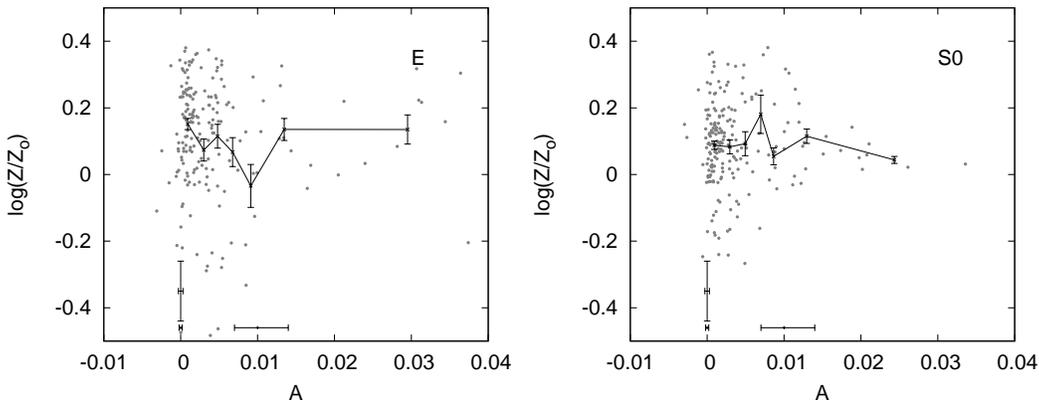}
\caption{Metallicity $\log_{10}(Z/Z_{\odot})$ as a function of asymmetry. 
Left panel: Es, right panel: S0s. 
Symbols, lines and error bars are as in Fig.~\ref{fig:asymm-color}.
\label{fig:asymm-met} 
}
\end{center}
\end{figure*}

One of the unavoidable limitations of the work of 
\citet{ss92} was that it was impossible to explore
quantitatively whether it was age or metallicity that
was driving the color--fine structure relation.  
Here, we make use of estimates of luminosity-weighted
stellar age and metallicity, available 
for almost all of the $\sim 2/3$ of our sample 
with SDSS spectra \citep{gallazzi05}, to address this question quantitatively.

Figs.~\ref{fig:asymm-age} and \ref{fig:asymm-met} 
show the run of luminosity-weighted age and 
metallicity as a function of asymmetry separately for E and S0 galaxies
(the trends in median age/metallicity are very similar).
There is a very strong correlation between age and asymmetry for 
both E and S0 galaxies (combined E/S0 sample has an essentially zero
chance of being due to chance alone, from a Pearson's 
or Spearman's correlation 
coefficient), and a substantially weaker or no correlation for
 metallicity (with a 0.031\% chance of being from chance alone
according to a Pearson's correlation coefficient, and a 1/3 chance
of being from chance alone as measured by a Spearman's coefficient). 

It is worth discussing briefly the sources of uncertainty
in this result. The random uncertainties ($\sim 0.1$\,dex in both 
log age and metallicity) are of little importance in 
this particular analysis, as one is seeking relative trends
within a large and homogeneous dataset.  
The composite
datapoints (mean for subsamples of 20 or more galaxies, typically)
have small random uncertainties ($\la 0.02$ dex), and differences
of $\sim 0.15$\,dex in e.g., age between $A \sim 0$ and $A\sim 0.01$ 
are detected with great significance.  Systematic effects could be more
important.  For example, if there were systematic differences
in alpha element overabundance as a function of asymmetry to the 
tune of $\sim 0.3$\,dex in alpha element overabundance (a {\it very} large
effect), a trend in age of up to 0.05\,dex could be produced (smaller than 
the observed trend in age with asymmetry)
in conjunction with an 
offset in metallicity of similar amplitude but {\it opposite} sign 
(\S 2.4 of \citealp{gallazzi05}).  Such a tendency is not seen; it is 
likely that alpha element overabundance is not an important driver of 
the observed trend in age (and weak or no trend in metallicity).  
Furthermore, as we saw earlier, changes in the inner/outer radii used to define asymmetry also change the results very little.  
One is left to conclude that genuine differences in star formation 
history, either through offsets in the age of the bulk of stars in 
the systems, or through a higher incidence of significant recent (last few Gyr) star formation activity between symmetric and asymmetric early-type
galaxies is the main driver of the age--asymmetry (and therefore
color--asymmetry) trend.  

We can therefore conclude that
the color--structure correlation is, to the extent that we can 
tell, driven primarily by age effects; this {\it observation} 
verifies the fundamental 
{\it assumption} made by \citet{ss92}.  The correlation is actually 
substantially clearer in age--asymmetry space, as the contribution from random 
scatter in the metallicities to the color--asymmetry trend act to scatter 
out the color--asymmetry relation.  


\section{Discussion}
\label{sec:disc}

Motivated by the work of \citet{ss92}, 
we showed that the color
offset from the early-type galaxy CMR
correlates with (primarily tidally-induced) asymmetry, although 
the scatter in the correlation is large.  
We showed, for the first time, that this correlation is driven primarily 
by age effects.  
In this section, we will explore further the meaning and limitations
of these results.  We will present a quantitative model-based
discussion of the results, focusing on the bearing that these
results have on the hypothesis that one of the primary formation 
routes of early-type galaxies is through galaxy merging.
Then, we will wrap up with a discussion of the limitations of this
work, and scope for future improvements.

\subsection{A quantitative model of red sequence offsets}

Like \citet{ss92}, we show a relationship between color
offset from the red sequence and asymmetric structure of a galaxy.
In \citet{ss92}, it was argued that the `fine structure' being 
measured was a strong indication of a previous gas-rich galaxy 
merger. Thus, in that paper, they presented a model in which 
early-type galaxies were the product of gas-rich major merging, 
and they studied the time taken for major merger remnants to age
as a way of translating their color offsets
into timescales for early-type galaxy evolution. 
Their model had a number of free parameters: the two key 
parameters $\tau_2$ and $\epsilon$ controlled the 
duration and prominence of star formation after the merger event.
They found that these parameters were of great importance in setting 
the timescales for fading of the remnant.  In models with a large
value of $\tau_2$ the remnant reddened only slowly, reaching 
the red sequence only after many Gyr.  In models with short-lived
post-merger star formation, they found a rapid evolution 
onto the red sequence.

One of the motivations of their work was to address 
the `gap' between clear remnants 
from relatively recent mergers ($\la 1$\,Gyr ago, 
e.g., NGC 3921 or NGC 7252)
and `ancient' early-type galaxies.  \citet{ss92} quote
King's question to Toomre at the 1977 Yale conference\footnote{Neither 
of us were at the Yale conference.}: ``You showed
us 10 merging pairs and then asked us to look for, or at least
accept the existence of, 500 remnants
from so long ago that they no longer bear the `made by Toomre' 
label.  I would be much more impressed if you showed us the 20 
or 30 such systems in the box immediately adjacent in your histogram.
What do these merged pairs look like in their next few galactic years?''.
While \citet{ss92} argued that the blue, structured early-type
galaxies in their study were reasonable candidates to 
fill the `King gap', the considerable sensitivity 
of the evolution of the merger remnant to the properties
of the merger and star formation after the merger made it difficult
to reach a unique conclusion.

In this subsection, we revisit this issue with the benefit of 
more advanced stellar population models, and more critically, 
a more physically-motivated model for the color evolution of 
early-type galaxies.  We will argue that the systems explored
in this paper (and in \citealp{ss92}) fill the `King gap', but that owing 
to natural scatter in the merger history of early-type galaxies
and in the stellar populations of these newly-formed early-type galaxies, 
it is impossible to connect systems one-to-one with a given 
merger time with data of the type that we use here; 
King's question was ill-posed, and will be 
very difficult to answer unambiguously.

\subsubsection{Model ingredients}

Clearly, there is a huge parameter space that can be explored
when modeling the evolving early-type galaxy population, and arriving 
at a single unique picture that is clearly superior to the others
will be impossible (as clearly articulated and demonstrated by \citealp{ss92}).

We take an approach that is motivated by the last decade of 
research on the evolution of the early-type galaxy population.
There are now a number of surveys that have empirically
tracked the evolution of the stellar mass density in early-type 
galaxies, with the general result that the growth in stellar mass
density is driven primarily by an increase in the {\it number} 
of early-type galaxies \citep{chen03,bell04,brown07,faber07}.  
Star formation is shut off 
in blue star-forming galaxies, and the remnant fades and reddens 
onto the red sequence \citep{bell07}.  
The exact mechanisms that shut
off star formation are still a matter of
some debate (merging and AGN feedback, environment, or gas exhaustion are
all possibilities; furthermore, the balance of such processes and timescales on which they operate is likely
to depend on a variety of parameters such as galaxy mass, or merger mass ratio; e.g., \citealp{johansson09feedback}); 
also open is the issue of how important
any burst of star formation might be before the truncation 
of star formation (see \citealp{robaina09} for a discussion 
of the average effects of galaxy mergers on star formation 
rate).  Notwithstanding this uncertainty about mechanisms 
for shutting off star formation, it is not unfair to assume that the timescale
for truncation of star formation is short, $\ll 1$\,Gyr\footnote{We 
are assuming in the terminology of \citet{ss92} that 
$\tau_2 \ll 1$\,Gyr.}.
Evidence indicating the short timescale for star formation 
truncation include the detection of rapid, arguably AGN-driven
winds in post-starburst\footnote{Galaxies with signatures
of no current star formation but substantial star
formation $<$1\,Gyr ago.} galaxies \citep{tremonti08}, 
the pronounced bimodality\footnote{such bimodality is 
washed out if transitions from blue to red
take any longer than 1\,Gyr} in the star formation rates of 
galaxies \citep{strateva01,bell04,sch08}, the 
high incidence of concentrated, post-starburst AGN galaxies
in the `valley' between the star-forming and
 non-star-forming galaxies \citep{schawinski07}, the modest
population of blue morphologically early-type galaxies
(if extended periods of star formation followed
a merger, a large fraction of star-forming early-type
galaxies would be observed; such a population is not 
common; \citealp{boris}, \citealp{ruhland07}, this paper)
and analysis of 
the spectra of early-type galaxies \citep{gallazzi06}.  
In the context of this paper (which was motivated originally by the merger
hypothesis), we will suppose that merging rearranges the stellar 
content of the remnant into a more spheroidal configuration and leads
to the suppression of future star formation through AGN feedback or 
some other mechanism.

Thus, we model a constantly-growing early-type galaxy population 
(an exercise similar in spirit to, e.g., \citealp{harker06}, 
\citealp{ruhland07}, or \citealp{naab09})
as a population of galaxies with 
constant star formation rate, that then stop forming 
stars at times drawn from the following probability 
distribution: $P(t) \propto 1 - t({\rm Gyr})/21$, where $t$ is the 
age of the galaxy in Gyr.  For concreteness, we choose to assign
$t_0 = 12$\,Gyr, i.e., the formation of early-type galaxies in 
this model started 12\,Gyr ago and continues at a slowly-reducing 
rate to the present day.  Our default models are 
truncation-only for simplicity; we show later that bursts
of star formation on truncation affect the results very little.
We assign an early-type galaxy a metallicity
drawn from a distribution with $\sigma_{\rm [Fe/H]} = 0.1$ and a mean 
of solar metallicity\footnote{The mean metallicity is 
of no importance in this paper, because we are exploring 
{\it residuals} from the color--magnitude relation.}, motivated
by the metallicity scatter inferred by \cite{gallazzi06}. 
We use the multi-metallicity PEGASE stellar population model to explore
the color evolution of the early-type galaxy population.
We neglect, for the sake of simplicity, possible differences in 
formation history as a function of galaxy mass\footnote{We see 
no significant difference in the behavior of this sample, 
when splitting into a high-mass
and low-mass subsample, justifying this oversimplification.}, while noting 
that the mean mass of our sample is somewhat less than $10^{11} M_{\odot}$.

\subsubsection{Major merger model}

\begin{figure*}
\begin{center}
\includegraphics[width=12.0cm]{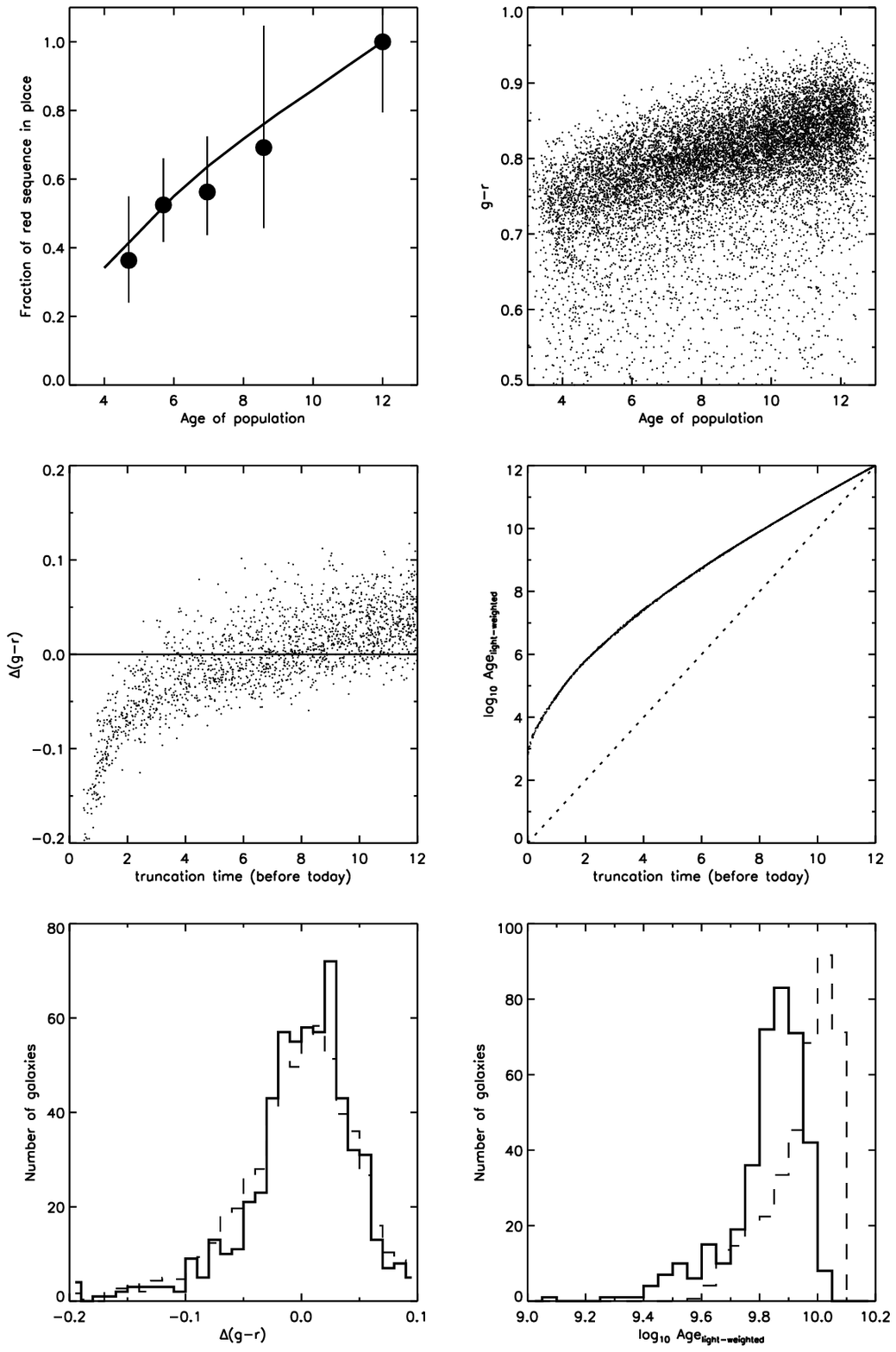}
\caption{A model in which major mergers create the early-type
galaxy population.  The top left panel shows the creation 
rate of early-type (red, in this case) galaxies as a function of
redshift from the models (solid line), and as measured by 
\cite{borch06}.  The top left panel shows the evolution 
of the predicted evolution of the color distribution of such a
population as a function of the age of the population (the
range $4-12$\,Gyr corresponds roughly to 
the interval $1>z>0$).  The middle left panel shows
the relationship between color offset $\Delta (g-r)$ and the 
truncation time, and the middle right panel between luminosity-weighted
age and truncation time.  The bottom panels show the predicted (dashed lines)
distributions of color offset and luminosity-weighted ages, and the 
observations (solid lines).
It is clear that the present-day color distribution of 
early-type galaxies is reproduced by a model in which 
the early-type galaxy population is being built up at the 
observed rate, and that in such a picture there is a 
broad but scattered correlation between merger$=$truncation time
and color offset from the color--magnitude relation.
\label{fig:maj} 
}
\end{center}
\end{figure*}

\begin{figure*}
\begin{center}
\includegraphics[width=12.0cm]{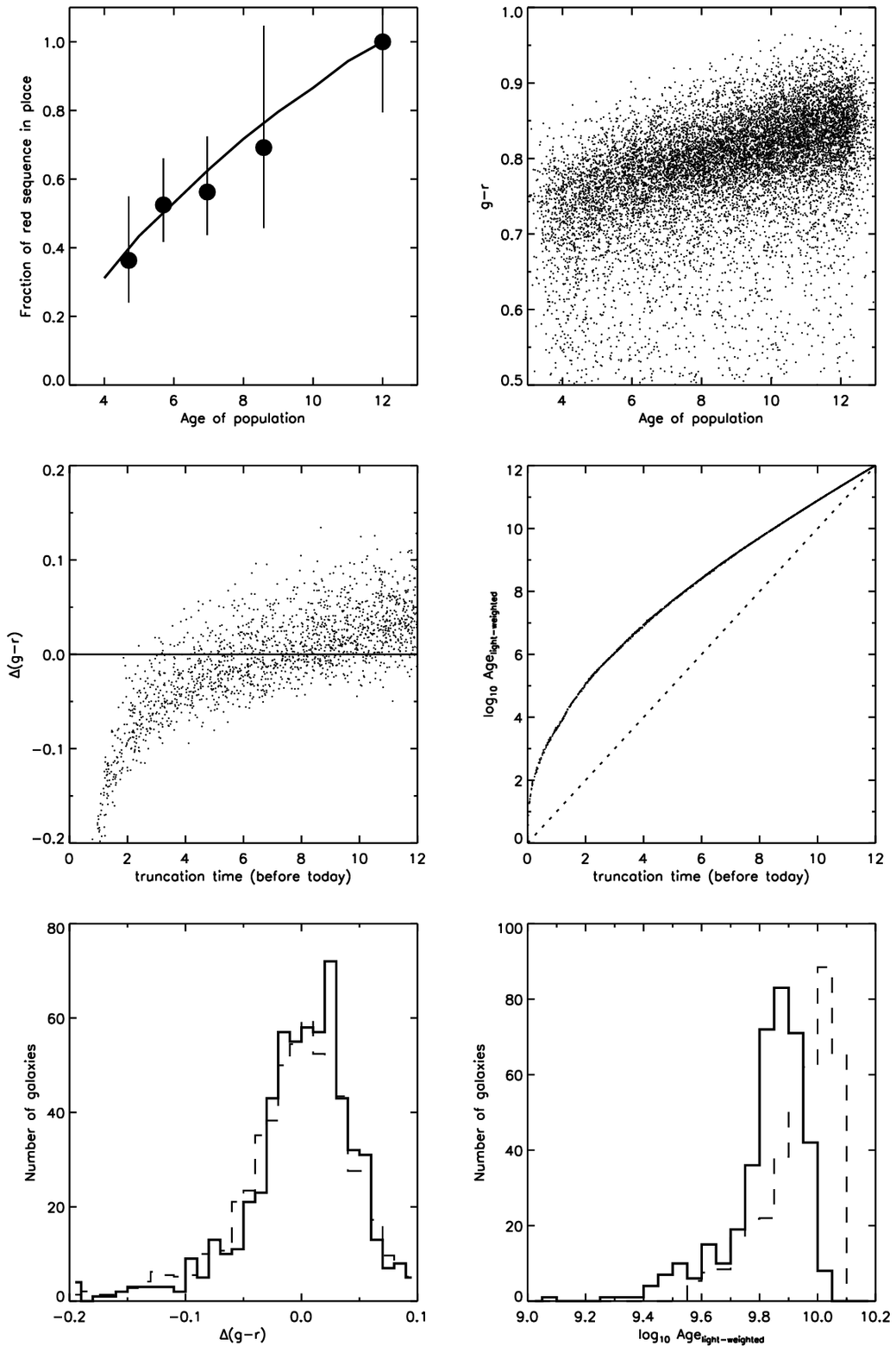}
\caption{A major merger model, but with a 10\% burst of 
star formation just prior to the truncation of star formation.
Our modeling conclusions are robust to a burst of star
formation on the final merger of a galaxy.
\label{fig:burst} 
}
\end{center}
\end{figure*}

The above model incorporates the effect of only one major merger: the 
one that created the early-type galaxy.  The results of 
such a model are shown in Fig.\ \ref{fig:maj}; 
the results of a very similar model where 10\% of 
the stars are formed in a burst on merging (motivated by 
\citealp{robaina09}) are
shown in Fig.\ \ref{fig:burst}, and are very similar
to the truncation-only model.   The upper
left panel shows the build-up in the number density of 
model red sequence, early-type galaxies, compared with 
observations of the build-up of the red sequence stellar
mass density from \citet{borch06}.  The upper right panel
shows the predicted evolution in the CMR 
zero point and scatter (think of this plot as a slice
through the evolving color--magnitude relation at constant stellar mass):
it is clear that there is a `core' of red sequence galaxies, with a 
tail of blue early-type galaxies.  

In the center left panel, 
we explore the relationship between truncation time (when star formation 
ceased) and 
color offset from the $z=0$, $t=$12\,Gyr red sequence. 
One can see that after truncation of star formation, the model
galaxy reddens rapidly, within $\sim 1$\,Gyr, to within 0.1 mag
of the red sequence.  The subsequent reddening of the population is 
much slower.  `Typical' structured early-type galaxies, with 
$\Delta (g-r) \sim -0.04$, would be interpreted to be between 
2 and 8 Gyr from their truncation event (depending on metallicity)\footnote{
Recall that many of the early-type 
galaxies with slightly blue colors
have little asymmetry in Fig.\ \ref{fig:asymm-color}, 
and the large magnitude
of this metallicity-dependent scatter in truncation time for 
a given color offset presents a natural and appealing explanation 
for this phenomenology. }.  It is to be noted that 
this timescale (in its value and scatter) is in rough agreement with 
\citet{ss92}; early-type 
galaxies less than $2\sigma$ from the 
locus of the color--magnitude are expected to be more 
than 1\,Gyr from their `creation event'.
The oldest red sequence galaxies
are {\it redder} than the core of the CMR.  This
is one of the reasons that the locus of the CMR 
appears to evolve {\it slower} than an ageing stellar population 
in such models; the average color of an early-type galaxy is always being 
pulled bluewards by new arrivals to the red sequence.   

We presented this relationship between truncation time and 
$\Delta (g-r)$, but can it have any relationship to reality (this, at some level, is the part of the question that \citealp{ss92} were less able to 
address quantitatively)?  Put differently, 
in a model which is constrained to reproduce the number density
evolution of early-type galaxies, is the color distribution of 
present-day early-type galaxies correctly predicted?  The result
of this exercise is shown in the lower-left panel of Fig.\ \ref{fig:maj}.
The solid histogram shows the color offsets from the red sequence
of the data, and the dashed histogram the major merger model.  
The two distributions are clearly very similar: a KS test yields
a $>10\%$ chance of the two distributions being drawn from the same 
distribution, which is a remarkable achievement for what is 
essentially a blind prediction, using only the observed 
red sequence number density growth and an assumed metallicity scatter.

In the center and lower right-hand panels, we show light-weighted
ages of the systems (comparable to the ages derived by \citealp{gallazzi05})
as a function of truncation time (central panel), and the histogram 
of light-weighted ages (lower panel).  It is clear that the luminosity-weighted
ages are skewed considerably towards older ages than the truncation times, 
as is expected from any model galaxies that have a constant star formation 
rate before the truncation of star formation.  The distribution of ages is 
reasonably encouraging, and has the asymmetry characteristic of the data.
The average age of the model is offset from the data; inasmuch as luminosity-weighted ages are considerably more model-dependent than color this offset could be
relatively easily remedied (but is not here, for the sake of simplicity).
For example, one could bring the start of star formation to $\sim 10$\,Gyr in the past and adopt a different parameterization of the early stages of truncation of star formation in early-type galaxy progenitors. 

\subsubsection{Minor merger model}

\begin{figure*}
\begin{center}
\includegraphics[width=12.0cm]{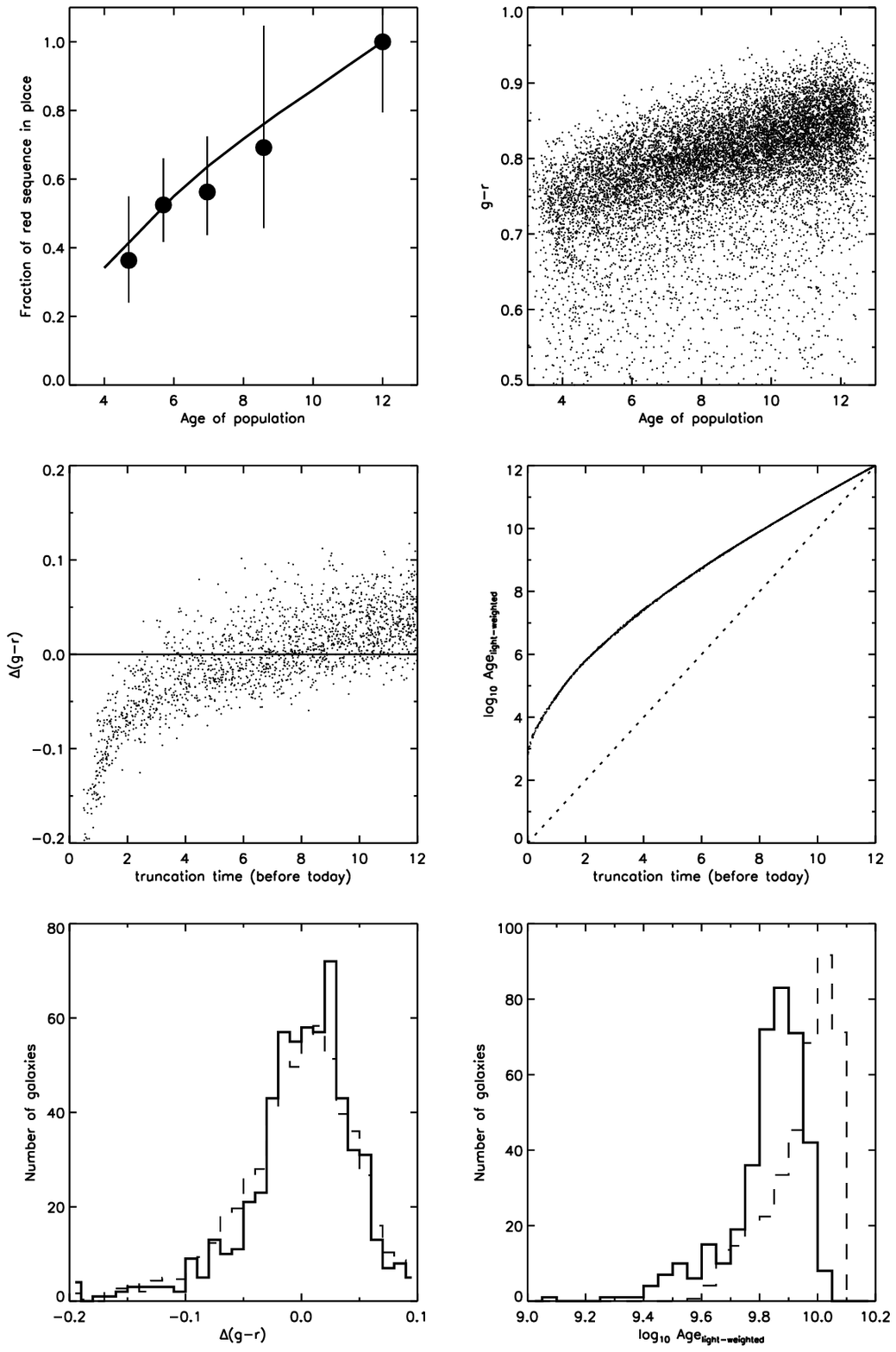}
\caption{A model in which major mergers create the early-type
galaxy population, with the addition of minor mergers.  
The figure is formatted similarly
to Fig.\ \ref{fig:maj}, except that in this case the middle
panels show population parameters as a function of the 
truncation time of the last event to take place, be it a major or 
a minor merger.
Again, it is clear that the color distribution of early-type
galaxies is well-reproduced by such a model, but that the 
addition of minor mergers scatters further the already scattered
correlation between color offset from the CMR 
and last truncation time.
\label{fig:min} 
}
\end{center}
\end{figure*}

Of course, minor mergers are much more frequent than major mergers, 
and are increasingly thought to be an important feature of the evolution 
of the early-type galaxy population (see, e.g., \citealp{naab09minor}, 
\citealp{bezanson09}, \citealp{hopkins10minor}).
The primary relevant effect here is that the 
minor merger of a gas-rich satellite onto a pre-existing 
early-type galaxy may drive the galaxy's color bluewards for a short time
and induce asymmetry.  In Fig.\ \ref{fig:min}, we show
a model that is identical to the major merger
model above, but with the early-type
galaxy population undergoing the accretion of a star-forming 
galaxy in the last 8 Gyr.  The probability of accretion 
is modelled as being constant over the last 8 Gyr, with 
a total probability of 1 that a given galaxy accretes
a star-forming satellite over that period.  
The mass ratio is randomly chosen from a uniform
distribution over the interval [0.0,0.4]; the metallicity of the 
satellite is assumed to be identical to the primary galaxy for 
modeling convenience.  

Fig.\ \ref{fig:min} shows many of the same phenomenologies as 
were seen in the major merger case (Fig.\ \ref{fig:maj}).  
The evolution of number of red sequence galaxies is similar, as is the 
color distribution produced at $t_0 = 12$\,Gyr.  The distribution of 
light-weighted ages is improved somewhat by the contribution of a 
small number of (bright) younger stars.  Furthermore, 
the relationship between $\Delta (g-r)$ and the `last' truncation time
(the time of the last event to take place, be it a major or minor merger)
is skewed considerably
towards younger ages; it is already possible to be very close
to the red sequence locus only $\sim 1$\,Gyr after a minor merger.

\subsubsection{Bringing it together: our interpretation of the asymmetry-color
relation}  \label{interp}

\begin{figure}
\begin{center}
\epsscale{1.0}
\plotone{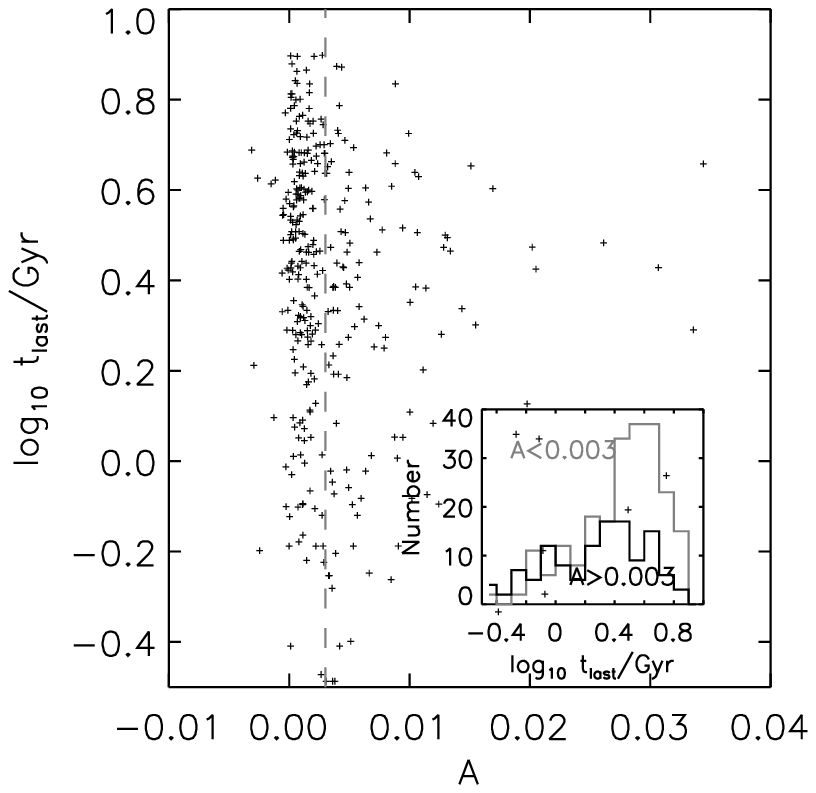}
\caption{Time of last major or minor merger, inferred statistically from the
luminosity-weighted age (from \citealp{gallazzi05}), 
as a function of asymmetry.  The dashed
line shows the dividing line between `symmetric' galaxies with $A<0.003$ and
`structured' galaxies with $A>0.003$.  The inset panel shows the histograms
of inferred last merger times for the symmetric and structured samples.
\label{fig:trunctime} 
}
\end{center}
\end{figure}

Thus, the two key conclusions of the modeling effort are:
{\it i)} the full color distribution of early-type galaxies 
at the present day is consistent with being built-up through 
truncation of star formation (as part of a merger event) 
at the rate inferred through observations of the build-up of the
red sequence, and {\it ii)} minor mergers can
lead to considerable short-lived color offsets (and presumably
asymmetries), complicating considerably their interpretation.

Despite the substantial scatter one expects in a major$+$minor
merger model, the correlations shown in the central panels
of Fig.\ \ref{fig:min} are still substantially stronger than 
the observed correlations between color/age and asymmetry
(although the correlations of \citealp{tal09} are almost
as strong as those in the central panels
of Fig.\ \ref{fig:min}; it is possible that for elliptical
galaxies residuals measured using deep data are well-correlated with 
time since last interaction). 
We quantitatively illustrate this issue in Fig.\ \ref{fig:trunctime}. 
We show the quantity $t_{\rm last}$, the lookback
time at which the last major or minor merger occurred in 
the major$+$minor merger model, statistically inferred from luminosity-weighted
age \citep{gallazzi05}.  
For each galaxy, a model galaxy with very similar 
luminosity-weighted age is found, and its $t_{\rm last}$ is assigned 
to the galaxy of interest\footnote{By doing this, we end up with more
scatter than is necessary in the final $t_{\rm last}$--asymmetry relation.
We choose a model galaxy with a given luminosity-weighted age at random, whereas
nature is likely to ensure that highly structured galaxies tend
to have a lower than average $t_{\rm last}$.  }. 
Such an inferred $t_{\rm last}$ is purely statistical, by design 
reproducing the trend and scatter in the middle right panel
of Fig.\ \ref{fig:min}; such a statistical
estimate is for illustrative purposes only and does not 
represent an estimate of $t_{\rm last}$ that would be accurate
in a galaxy-by-galaxy sense. 
We show the $t_{\rm last}$ inferred in this 
way as a function of asymmetry; the inset panels show the $t_{\rm last}$
distribution inferred for symmetric ($A<0.003$) and structured ($A>0.003$)
galaxies.  
It is clear that asymmetry and $t_{\rm last}$ correlate
only very weakly; while structured galaxies have a weak tendency towards lower
$t_{\rm last}$, there are a number of structured galaxies with high 
$t_{\rm last}$ (i.e., rather red stellar populations).  Put differently, 
this figure attempts to encapsulate quantitatively the 
fact that {\it neither} the stellar population parameters
{\it nor} the asymmetry are perfect clocks: the scatter
in star formation history, gas content, mass ratio and details of the 
merger orbits appear to add very significant scatter
to the relationship between the stellar population and structural `clocks'.

\begin{figure*}
\begin{center}
\includegraphics[width=14.0cm]{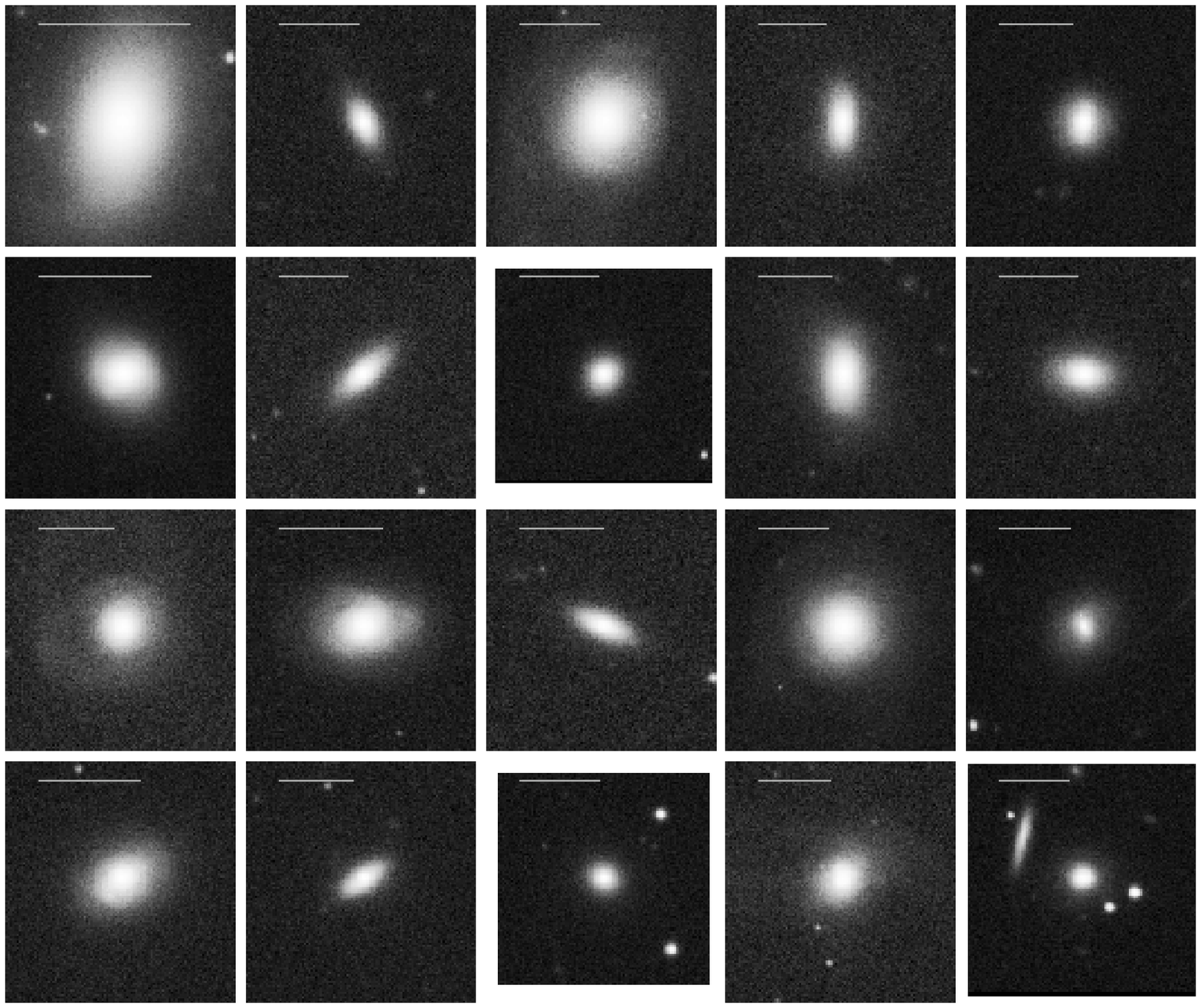}
\caption{The 20 bluest E/S0 galaxies. 
Each box is approximately $55\times55$ arcsec. A 10kpc horizontal line is shown 
for each galaxy.  
\label{fig:blue} 
}
\end{center}
\end{figure*}

\begin{table*}
\caption{Parameters of the 20 bluest E/S0 galaxies \label{tab:blue}}
\begin{center}
\begin{tabular}{c c c c c c}            
\hline
\hline
  RA &  Dec &  $\Delta (g-r)$ &  log (age [Gyr]) &  log (Z/Z$_{\odot}$) &  A \\
\hline
 224.1003 & 49.6960 &   -0.14 &  \nodata & \nodata &  0.002 \\
 166.3744 & 3.6600 &   -0.14 &  9.56 & -0.28 &  0.003 \\
 142.5555 & 49.4882 &   -0.14 &  \nodata & \nodata &  0.005 \\
 227.0540 & 56.4228 &   -0.14 &  9.32 & -0.03 &  0.012 \\
 169.7402 & 58.0566 &   -0.15 &  \nodata & \nodata &  0.003 \\
 17.9440 & -0.6645 &   -0.15 &  \nodata & \nodata &  0.002 \\
 226.6311 & 40.6962 &   -0.16 &  \nodata & \nodata &  0.013 \\
 202.2038 & 53.4430 &   -0.18 &  \nodata & \nodata &  0.009 \\
 135.1501 & 46.6863 &   -0.19 &  \nodata & \nodata &  0.007 \\
 167.6620 & 3.7555 &   -0.19 &  9.51 & -0.46 &  0.005 \\
 32.8899 & 13.9171 &   -0.19 &  8.84 & -0.04 &  0.016 \\
 11.2468 & -8.8897 &   -0.20 &  9.49 & 0.03 &  0.024 \\
 50.8886 & -0.4385 &   -0.20 &  \nodata & \nodata &  0.002 \\
 135.7590 & 40.4340 &   -0.21 &  \nodata & \nodata &  0.012 \\
 153.9239 & 7.0522 &   -0.21 &  \nodata & \nodata &  0.013 \\
 199.9037 & 3.0327 &   -0.22 &  8.87 & -0.20 &  0.037 \\
 129.9133 & 3.8285 &   -0.24 &  \nodata & \nodata &  0.022 \\
 210.1692 & -1.9217 &   -0.29 & \nodata  & \nodata &  0.004 \\
 120.8668 & 25.1026 &   -0.31 &  \nodata & \nodata &  0.039 \\
 205.3000 & 1.7798 &   -0.42 &  9.39 & -1.08 &  0.013 \\
\hline
\end{tabular}
\end{center}
\end{table*}

In this kind of picture, while one
expects a broad correlation between color offset from the red sequence 
and asymmetry (as is observed), there will be a lot of scatter in this 
relationship for astrophysical reasons.  
Some relatively recent minor accretions, or 
interactions between already non-star-forming
progenitors (dry mergers; \citealp{vd05,bell06,dmac08}), might be richly
structured but already red; whereas some more ancient major interactions 
may remain blue for much longer, showing little asymmetry but blue colors.
In such a picture, finding galaxies to fill the `King gap' is 
not a particularly well-posed exercise.  It is of course
possible to define samples that will likely contain some 
fraction of the most recently-formed early-type galaxies.
Such a candidate sample is presented in  Fig.~\ref{fig:blue} 
and Table~\ref{tab:blue}, where we present 
the 20 bluest objects in our sample, with  $\Delta (g-r) < -0.14$. 
Inspection of Fig.\ \ref{fig:blue} shows that many of them clearly
show asymmetries generated by tidal effects; yet, some of this sample 
are not obviously asymmetric, illustrating the challenges 
of uniquely identifying a sample of `young' early-type galaxies.

Yet, we would argue that 
there should be little concern about this difficulty to 
uniquely identify intermediate-age merger remnants.  Much more important
in our view is the finding that in this model, 
in which early-type galaxies are being produced at the right rate 
by major interactions (with or without the additional accretion of satellites),
the correct distribution of early-type galaxy colors is straightforwardly 
reproduced.   This indicates that as an ensemble, the (stellar population) 
properties
of the early-type galaxy population are in accord with a model 
in which they are being built up by merging at the observed rate.  Our 
result confirms a broad link between asymmetric 
(largely tidally-induced) structure 
and stellar population parameters, confirming a qualitative prediction
of the merger hypothesis.  Detailed $N$-body and hydrodynamical modeling
will be necessary to test this link {\it quantitatively}, by 
providing directly the predicted distribution of 
galaxies in the color/age vs.\ asymmetry plane as predicted 
by assembly histories derived from simulations of galaxy 
formation in a cosmological context.

\subsection{Musings on the role of dissipationless (dry) merging}
\label{sec:dry}

\begin{figure*}
\begin{center}
\includegraphics[width=14.0cm]{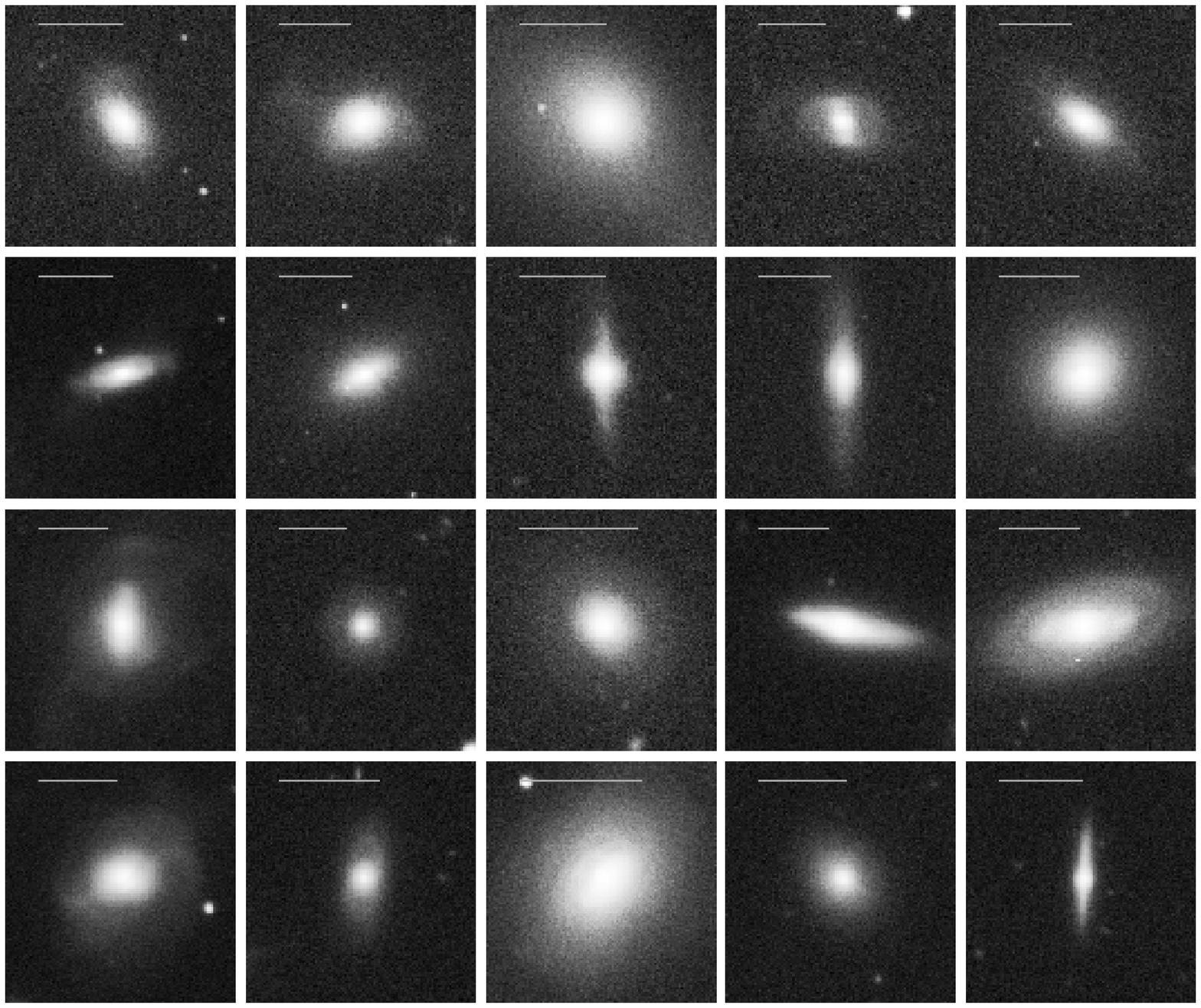}
\caption{A sample of 20 red, asymmetric E/S0 galaxies. 
Each box is approximately $55\times55$ arcsec. A 10kpc horizontal line shown 
for each galaxy.  
\label{fig:red} 
}
\end{center}
\end{figure*}

In recent years, it has been demonstrated that 
largely dissipationless merging between already-formed
early-type galaxies (dry mergers) play a significant
role in the build-up of the early-type galaxy population
\citep{bell04,vd05,bell06,lotz06,faber07,skelton09}.  In particular, 
the most massive early-type galaxies appear to 
grow {\it only} by dry merging
\citep{bell04,faber07,dmac08}, although what that
growth rate is remains somewhat unclear
\citep{vd05,bell06,masjedi06,scarlata,dmac08}.

In this context, it is of interest to note the population 
of relatively red ($\Delta(g-r) \sim 0$) and richly
structured E and S0 galaxies in Figs.\ \ref{fig:asymm-color}
and \ref{fig:asymm-age}.  
We show a set of 20 randomly selected 
red highly asymmetric galaxies with  $\Delta(g-r)>-0.1$ and $A>0.02$ 
in Fig.~\ref{fig:red}.
These systems are clearly undergoing
(or have undergone) interactions, both minor and major, with 
already early-type galaxies.
The relative prominence of this tail of red asymmetric E/S0 
galaxies appears to be relatively independent of luminosity\footnote{This
was checked
by splitting the sample in half by luminosity.  There was no significant 
difference between the red asymmetric early-type galaxy fractions of 
the bright and faint bins.}
(somewhat in contrast to the naive expectation that dry interactions/mergers
are more important for more luminous systems), indicating that 
such interactions are possible for systems of a wide range of 
luminosities from $L^*/5$ right up to the most luminous systems 
probed in our study $\sim 2-3 L^*$.  

It is of interest that many of 
the most structured early-type galaxies have this signature.  
Without a wide range of $N$-body/hydrodynamical simulations 
to guide us, it is difficult to properly interpret this behavior.  
Two aspects will clearly play a role in setting the
relative prominence of blue vs.\ red richly-structured 
early-type galaxies: the ratio of blue vs.\ red
merger partners typical of the mergers that create
early-type galaxies, and the issue of when one classifies 
a merger remnant as an early-type galaxy (almost always for 
two interacting early-type galaxies, as opposed to only at very late times for
the remnant of a merger between two gas-rich galaxies).  

It is also of historical interest to ask why there were relatively few such
systems in the work of \citet{ss92}.  Number statistics clearly 
will play a role, but also important is the differing nature
of our measures of tidal structure in early-type galaxies.
\citet{ss92} attempted to measure {\it fine} structure 
characteristic of mergers between gas-rich systems, whereas
we explore asymmetry, which is also sensitive to large-scale
asymmetries and debris fields, more characteristic of 
mergers between early-type galaxies.  This aspect, along with 
details of initial sample selection, is likely to play a role in 
this small difference between our two studies.

\subsection{Comparison with Tal et al. (2009)}

\citet{tal09} choose a complementary method to explore this 
issue, by quantifying the residuals from a smooth model
fit for a sample of elliptical galaxies and correlating them 
with color offset from the CMR (residuals from a smooth 
fit for S0s would reveal the disk/bulge transition and bars, and
would be significantly less meaningful).  They used very deep 
data for 55 systems and traced tidal structure out to large radius, where
dynamical times are long.  They found that a large fraction of systems 
were disturbed, and found a clear but scattered correlation 
between color offset from the CMR and residuals, as seen by 
\citet{ss92}.  Their results can be regarded as a high sensitivity
analysis of the main body of the correlation shown in 
Fig.\ \ref{fig:asymm-color}, where their higher sensitivity
to faint debris has brought the color-residual correlation into 
sharp focus.  Our result complements this analysis by i) using 
a larger sample and including S0 galaxies, showing it to be 
generally applicable to the population as a whole (albeit with 
a lower sensitivity measure of structure, largely because of 
limitations in SDSS depth), and ii) using age measures
to explicitly illustrate that the main driver of such 
behavior is an age--structure relation.

\subsection{Limitations of this approach, and outlook}

As alluded to in the previous sections, a significant 
(and largely unavoidable) limitation of our study is the choice
of asymmetry as a metric of tidally-induced disturbances.
Our choice of asymmetry was motivated by the overriding consideration
that the measurement be algorithmic and reproducible; asymmetry
was chosen over other reproducible descriptions of structure 
(such as residuals from a smooth model fit; e.g., \citealp{dmac02} and
\citealp{tal09}) 
to avoid symmetric bars or rings from 
contributing to our measure of structure.  Yet, a key limitation 
of our metric is that systems can be asymmetric for a variety of 
reasons: asymmetry may be produced by fly-bys or mergers, major or minor
interactions,  ongoing or past interactions, gas-rich or gas-poor mergers.
As such, asymmetry is a valuable but ultimately blunt tool; 
in this respect, our work is a 
useful way to start attacking the problem, but is not 
capable of answering higher-level, more detailed questions about 
the origin of the early-type galaxy population.

In principle, one could attempt to construct
a richer description of structure, which would be 
more clearly correlated with particular types of interaction
(e.g., a measure of the spatial scale of tidal debris, or 
a measure of the luminosity in tidal debris).  Calibration 
with simulations would be a critical step in the establishment of
such metrics (indeed, we have not yet calibrated our measure 
of asymmetry as part of this work).  Yet, in 
practice, such metrics have proven difficult to reliably construct; 
the fine structure parameter of \citet{ss92} is a good example of
such a metric, which may be more physically meaningful but 
is difficult to reproduce.

\section{Conclusions}
\label{sec:concl}

One of the key predictions of the merger hypothesis for the 
origin of early-type galaxies is that tidally-induced asymmetric structure
should correlate, at least in a broad sense, with signatures of 
a relatively young stellar population.  Motivated by this 
argument, \citet{ss92} studied the relationship between color
offset from the CMR and 
tidally-induced fine structure, finding a correlation 
between the two quantities.  They modeled the colors of these structured
early-type galaxies, finding that such colors can be reproduced by 
a variety of models in which star formation truncates reasonably 
quickly, with typical timescales of a few Gyr to reach the colors
characteristic of ancient early-type galaxies (although with much 
model dependence).

In this paper, we have re-examined this issue, incorporating 
a number of improvements over the work of 
\citet{ss92}: a sample from 2MASS/NED and the SDSS
that is nearly ten times larger with well-measured
colors; the use of a non-ideal, but reproducible and meaningful metric
for tidally-induced structure, asymmetry; the use of luminosity-weighted
ages and metallicities from the work of \citet{gallazzi05} using the 
SDSS; and, the ability to construct a more physically-motivated model 
of early-type galaxy evolution with which to interpret the results.

We found, in agreement with \citet{ss92} and \citet{tal09}, 
a correlation between 
offset from the CMR and asymmetry.  Inspection 
of asymmetric systems gave weight to the notion that the main driver
of asymmetries was galaxy interactions (although a variety of 
interactions can create asymmetries, leading to a natural ambiguity when 
interpreting the results).  We demonstrated, for the first time, 
that age effects are driving this correlation.  A fraction 
of asymmetric early-type galaxies have normal colors/ages, 
characteristic of mergers between already-formed 
early-type galaxies (dry mergers).
The empirical 
correlation between stellar population age and tidally-induced
asymmetries, for the bulk of the population, 
is the key result of this paper, and is consistent with 
 the basic prediction of the origin of early-type galaxies 
through galaxy merging, and their modification by accretion 
of gas-rich satellites.

We interpreted these results in the context of a model in which 
the number of early-type galaxies is increasing constantly
with time, in quantitative agreement with measurements of 
the evolving number density of red sequence (early-type) galaxies.
Assuming that the ultimate effect of a galaxy merger is to 
truncate star formation (through either gas consumption or 
AGN feedback), we find that such a model reproduces
the distribution of color offsets from the CMR; 
put differently, the merger hypothesis appears to give
the correct distribution of present-day early-type galaxy colors.
Under the assumption that both light-weighted stellar age and 
asymmetry are good `clocks' of the time elapsed since the last
major or minor merger, such a model predicts
a rather tighter correlation between age and asymmetry than is 
observed.  This indicates that astrophysical sources of
scatter, e.g., the star formation histories of the progenitors, 
their gas content, and mass ratios, play an important role
in setting the properties of the remnants.  Thus, the broad
correlation between age and asymmetry is in qualitative (but not 
yet quantitative)
agreement with such a model.  The existence of 
such a large scatter precludes the unambiguous identification 
of `young' early-type galaxies from stellar population and asymmetry 
data alone; confirming the merger hypothesis using a few 
individual examples will be a challenging and ambiguous exercise.

We conclude that the properties of the early-type
galaxy population as a whole are consistent with the bulk of
them being formed by galaxy merging (some of these mergers
are between already-formed early-type galaxies).  In the case of 
stellar population parameters, there is a quantitative 
match between toy models of merger-driven growth and the observations.
In the case of asymmetries, the existence of a broad and scattered
correlation between stellar population parameters and asymmetry 
is in qualitative agreement with the expectations from the merger
hypothesis, but our toy models (and even more sophisticated models) 
are currently incapable of providing expectations for how asymmetry
should vary with time and interaction phase.  Further work, 
using simulations to predict and calibrate the distributions of 
asymmetries, and using more sophisticated descriptors of galaxy 
structure, will help to more deeply explore this issue in the next years.

\acknowledgements
We thank the referee for a thorough and helpful report, 
Ignacio Trujillo for image simulations used to verify the 
image fitting, and Marco Barden for assistance with GALFIT.  We thank
David Hogg for useful suggestions.
E.\ F.\ B.\ thanks the Deutsche Forschungsgemeinschaft for their support 
through the Emmy Noether Program.  
This publication makes use of data products from the 
{\it Two Micron All Sky Survey}, which is a joint project of the 
University of Massachusetts and the Infrared Processing and 
Analysis Center/California Institute of Technology, funded by 
the National Aeronautics and Space Administration and 
the National Science Foundation.
This publication also makes use of the {\it Sloan Digital
Sky Survey} (SDSS).
Funding for the creation and distribution of the SDSS 
Archive has been provided by the Alfred P.\ Sloan Foundation, the 
Participating Institutions, the National Aeronautics and Space Administration,
the National Science Foundation, the US Department of Energy, 
the Japanese Monbukagakusho, and the Max Planck Society.  The SDSS
Web site is \texttt {http://www.sdss.org/}.  The SDSS Participating
Institutions are the University of Chicago, Fermilab, the Institute 
for Advanced Study, the Japan Participation Group, the Johns Hopkins
University, the Max Planck Institut f\"ur Astronomie, the Max
Planck Institut f\"ur Astrophysik, New Mexico State University, 
Princeton University, the United States Naval Observatory, and 
the University of Washington. 
This publication also made use of NASA's Astrophysics Data System 
Bibliographic Services.

\end{document}